\documentclass[reprint,amsmath,amssymb, aps, prl]{revtex4-1}

\usepackage{graphicx}
\usepackage{epsfig}
\usepackage{color}
\usepackage{amsmath}
\usepackage{amsfonts}
\usepackage{mathrsfs}
\usepackage{txfonts}
\usepackage{color}

\usepackage{hyperref}

\begin{document}
	\title{Clean Floquet Time Crystals: Models and Realizations in Cold Atoms}
	\author{Biao Huang$^1$}
	\email{phys.huang.biao@gmail.com} 
	\author{Ying-Hai Wu$^2$}
	\author{W. Vincent Liu$^{1,3,4}$}
	\email{wvliu@pitt.edu}
	\affiliation{
		$^1$Department of Physics and Astronomy, University of Pittsburgh, Pittsburgh PA 15260, USA\\
		$ ^2 $Max-Planck-Institut f\"{u}r Quantenoptik, Hans-Kopfermann-Stra\ss e 1, 85748 Garching, Germany\\
		$ ^3 $ Wilczek Quantum Center, School of Physics and Astronomy and T. D. Lee Institute,  Shanghai Jiao Tong University, Shanghai 200240, China\\
		$ ^4 $Center for Cold Atom Physics, Chinese Academy of Sciences, Wuhan 430071, China
		}
	\begin{abstract}
		Time crystals, a phase showing spontaneous breaking of time-translation symmetry, has been an intriguing subject for systems far away from equilibrium. Recent experiments found such a phase both in the presence and absence of localization, while in theories localization by disorder is usually assumed a priori. In this work, we point out that time crystals can generally exist in systems without disorder. A series of clean quasi-one-dimensional models under Floquet driving are proposed to demonstrate this unexpected result in principle. Robust time crystalline orders are found in the strongly interacting regime along with the emergent integrals of motion in the dynamical system, which can be characterized by level statistics and the out-of-time-ordered correlators. We propose two cold atom experimental schemes to realize the clean Floquet time crystals, one by making use of dipolar gases and another by synthetic dimensions.
	\end{abstract}
	\maketitle

	\noindent{\em Introduction --- }The recent realizations of Floquet (or discrete) time crystals have drawn much attention \cite{sondhi2015, sondhi1, nayak, sondhi2, yaoashvin,  nayakprethermal, sondhi1612, dtcexpt1, dtcexpt2, sacha}. A common feature of these systems is that certain physical observable $ \hat{O} $ shows a rigid reduced periodicity $ \langle \hat{O}\rangle(t+nT)=\langle \hat{O} \rangle (t), n\ge 2 $, compared with the Floquet driving period $ T $ of the Hamiltonian $ H(t+T)=H(t) $. As originally conceptualized in Ref. \cite{wilczek0, wilczek1, wilczek2},  ``time-crystals'' are regarded as a new addition to the concept of spontaneous symmetry breaking, for the temporal translation symmetry missing for nearly a century.  
	
	Early discussions of time crystals \cite{wilczek1,wilczek2,earlyexample,earlyexample2} concluded with a no-go theorem \cite{oshikawa} forbidding such a phase in equilibrium. Consequently, a new generation of periodically-driven models were proposed \cite{sondhi2015, sondhi1, nayak, sondhi2, yaoashvin}, with results that challenge our understanding of dynamical interacting systems. Unlike the usual quasi-static examples such as charge pumping \cite{pumpingThouless,pumpingB,pumpingF} or lattice shaking \cite{shaking1,shaking2}, the Floquet time crystal lives in the regime with large driving amplitude and resonant frequencies, surprisingly robust against chaotic behaviors, such as in turbulence \cite{landau,zoran,jasoncascade}. It is therefore natural to ask what serves as the stabilizer against butterfly effects and heating.
	
    A key strategy in recent theories is to employ non-ergodic systems to resist trivialization of dynamics due to thermalization \cite{nayak,sondhi1,sondhi2,yaoashvin}. Besides the fine-tuned integrable Hamiltonians, many-body localized (MBL) systems consist of the most well-studied examples showing robust non-ergodicity. As such, it is assumed {\em a priori} in most theories that stable time-crystal phase can only occur in the MBL regime with strong spatial disorder \cite{nayak,yaoashvin,dtcexpt1}. However, a recent experiment on  nitrogen-vacancy (NV) centers performed by Choi {\em et. al.} demonstrated an alternative possibility \cite{dtcexpt2}, where time crystals formed regardless of the delocalization by the three-dimensional spin-dipolar interactions. It was also emphasized that the system is not in a pre-thermal regime \cite{nayakprethermal,dtcexpt2}. The experimental breakthrough indicates the tantalizing possibility of seeking for stable time-crystals without the aid of localization, and the theoretical need to understand the time-crystal phase in this regime.
	
	The purpose of this work is to demonstrate through a simple model that stable time crystals can exist in the strongly interacting regime {\em completely without disorder}. These Floquet-ladders we propose represent a large class of models including, as special cases, the quenched Ising chain \cite{nayak,yaoashvin,sondhi1,sondhi2} discussed before. Within certain parameter {\em regions}, the persisting double-periodic oscillation modulates with time spans that scale exponentially with system sizes. Unlike the ``MBL time crystal"  \cite{nayak,yaoashvin,sondhi1,sondhi2} which inherits integrability from a static MBL-Hamiltonian, these ``clean time crystals" exhibit emergent integrability through dynamics and is a property of the Floquet evolution operator. Such a character is illustrated by the level statistics and out-of-time-ordered-correlators (OTOC) in different parameter regimes.  Moreover, these phenomena even survive when the interactions are modified to those that can be readily realized in current cold atom experiments. The generality of our results clearly suggests an exciting field of studying time crystals in various clean systems with more intriguing properties.

	{\em Definition of time crystal---}Periodic motions exist widely in dynamical systems, ranging from Rabi oscillations \cite{qmtextbook} to Josephson effects \cite{josephson}  and Zitterbewegung \cite{zitter}. More generally, if one picks an arbitrary initial state, the unitary time-evolution $ e^{-iHt/\hbar} = \sum_n |n\rangle e^{-iE_nt/\hbar}\langle n| $ may fairly endow the evolved state certain oscillations. Therefore, restrictions must be applied to screen out some periodic motions that are already well-understood without involving a new name. Here we give a phenomenological definition of non-equilibrium time crystal by selecting oscillations that are emergent from many-body dynamics. Specifically, there should exist a physical observable $\hat{O}$ and a class of initial states $|\psi\rangle$, such that
	\begin{equation}
	f(t) = \lim_{L\rightarrow\infty}\langle \psi| \hat{O}(t) |\psi \rangle
	\end{equation}
	satisfy {\em all} of the three conditions:
	{\bf(A)} Time-translation-symmetry-breaking, which means $ f(t+T)\ne f(t) $ while the Hamiltonian has $ H(t+T)=H(t) $;
	{\bf(B)} Rigidity: $ f(t) $ shows a fixed oscillation frequency without fine-tuned Hamiltonian parameters.
	{\bf(C)} Persistence: the non-trivial oscillation with fixed frequency must persist to indefinitely-long time when first taking system size $L$ to the thermodynamic limit. 
	
	The above definition is inspired by making analogy to the familiar charge-density-wave (CDW). Condition (A) rules out oscillations trivially following the external drive, which functions as ``temporal lattice potentials''. The rigidity of frequency in condition (B) requires many-body origins, resembling the rigidity of wave-vector for density-modulation in CDW given by Fermi-surface nesting \cite{xiaohuangshu}. Condition (C) is added to distinguish a stable time crystal from a quasi-stable, i.e. a pre-thermal one \cite{nayakprethermal}, or accidental oscillations lasting for short periods. See also Ref. \cite{sondhi1, nayak, sondhi2, yaoashvin, sondhi1612} emphasizing different aspects of the definitions respectively.

	\begin{figure}[h] 
		\begin{flushleft}
			\parbox{1cm}{\flushleft \quad (a)}
			\parbox{6cm}{\includegraphics[width=6cm]{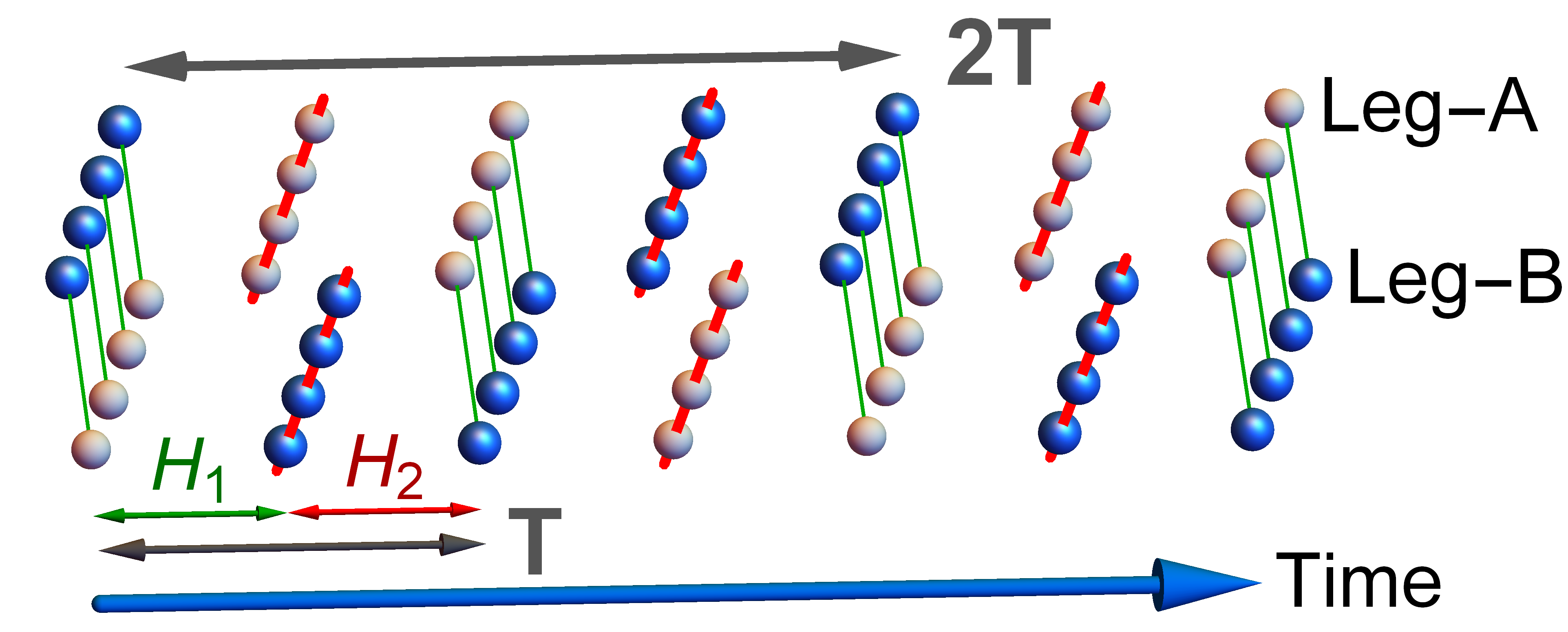}}\\
			$ \quad $\\ $ \quad $\\
			\parbox{1cm}{\flushleft \quad (b)}
			\parbox{6cm}{\includegraphics[width=5.5cm]{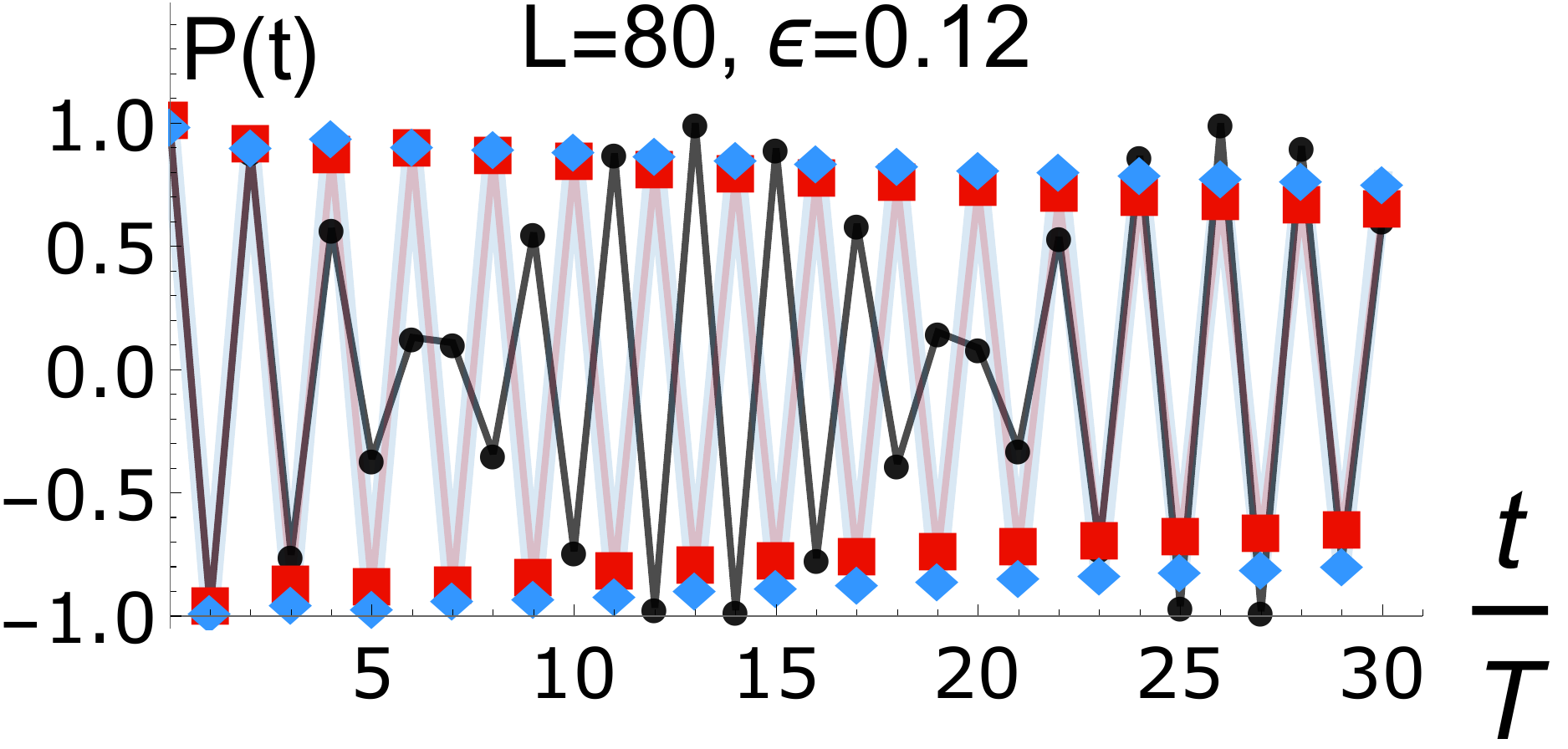}
			}
			\parbox{1cm}{\includegraphics[width=1.1cm]{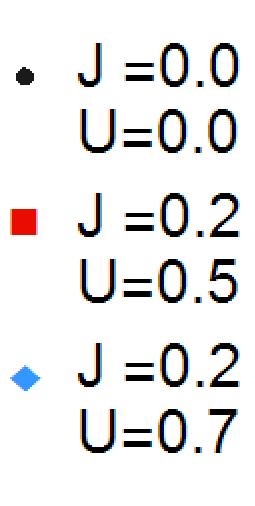}}
		\end{flushleft}
		\caption{\label{fig:twochains} (a) Schematic plot for the Floquet-ladder. Green and red lines indicate the inter-chain (Eq.~(\ref{h1})) and intra-chain couplings (Eq.~(\ref{h2})) respectively, which alternate during the binary drive. Blue dots represent occupied sites. In time crystal regime, density distribution in two chains shows rigid reduced periods $ 2T $.  (b) DMRG result for density polarization $ P(t) $ (Eq.~(\ref{pt})) under perturbation $ \varepsilon=0.12 $ at stroboscopic time (lines are guides to the eyes). The interaction $ U $ rigidifies the $ 2T $ periodicity, signifying a time-crystal phenomenon. Here the lattice size is $ L=80 $ for each chain, $ \Delta=0.1 $, and the open boundary condition is used.}
	\end{figure}
	{\em Model---}We introduce a clean Floquet-ladder model that turns out to satisfy all of the conditions (A)-(C). The Hamiltonian is under binary quench with periodicity $ T = t_1 + t_2 $, where during
	\begin{eqnarray}\label{h1}
	 t_1: && \quad H_1 = -J'\sum_{i=1}^L(a_i^\dagger b_i + b_i^\dagger a_i),\quad \frac{J' t_1}{\hbar} = \frac{\pi}{2}+\frac{t_1}{\hbar}\varepsilon,\\ \nonumber
	 t_2: && \quad H_2 = -J\sum_{i=1}^L (a_{i+1}^\dagger a_i +b_{i+1}^\dagger b_i + h.c.) \\ \label{h2}
	 && \qquad + U\sum_{i=1}^L (n^A_{i}n^A_{i+1} + n^B_{i} n^B_{i+1}) + \Delta\sum_{i=1}^L (n_i^A - n_i^B).
	\end{eqnarray}
	See Fig. \ref{fig:twochains}(a) for illustrations.
	Here $ a_i^\dagger $ ($ b_i^\dagger $) creates a particle in leg-A (-B), $ n_i^{A,B} = a_i^\dagger a_i $ (or $b^\dagger_i b_i$) is the particle number operator, and $ L $ is the number of sites in each leg.
	The evolution operator at stroboscopic time is
	\begin{equation}\label{uf}
	  U(nT) \equiv (U_F)^n = \left( e^{-iH_2t_2/\hbar} e^{-iH_1t_1\hbar} \right)^n
	\end{equation}
	where $ U_F $ is the Floquet operator. The physics is controlled by dimensionless parameters $ ( \varepsilon t_1, U t_2, J t_2, \Delta t_2)/\hbar $, which will be denoted simply as $ (\varepsilon, U, J, \Delta) $ later on.	To compare with previous works using an Ising chain \cite{nayak, yaoashvin, sondhi1, sondhi2}, we note that for either spinless fermions or hard-core bosons, our model maps to {\em two} coupled spin-$ 1/2 $ XXZ chains, and is therefore generically different (in additional to the lack of disorder) except in the special limit $ J=0 $ and $ n_i^A+n_i^B=1 $ \cite{supp}.

	The general characters of our model are as follows. Dynamics during $ t_1 $ resembles single-particle Rabi oscillations of particles between two chains $ U_1= e^{-iH_1t_1/\hbar} $, i.e. $ U_1^\dagger a_j^\dagger U_1 =  i \cos(\varepsilon)b_j^\dagger -\sin(\varepsilon)a_j^\dagger $, and $ U_1^\dagger b_j^\dagger U_1 =  i \cos(\varepsilon)a_j^\dagger -\sin(\varepsilon)b_j^\dagger $. During $ t_2 $, each chain is experiencing nearest-neighbor interactions separately. Define the physical observable as the density polarization $ P(t) $ between two chains,
	\begin{equation}\label{pt}
	P(t)=\frac{1}{L}\sum_i \langle \psi(t)| \hat{P}_i |\psi(t)\rangle, \qquad
	\hat{P}_i =  a_i^\dagger a_i - b_i^\dagger b_i.
	\end{equation}
	When $ \varepsilon=0 $, its periodicity is strictly $ 2T $ regardless of $ H_2 $. But the period of $ P(t) $ is unstable against perturbations $ \varepsilon $ to the ``Rabi frequency'': see the example of $ J=U=0 $ in Fig. \ref{fig:twochains}(b). The essential feature is that the dynamics during $ t_2 $, though keeping $ P(t) $ unchanged, functions as a many-body synchronizer for the $ 2T $ periodicity of $ P(t) $ and rigidifies the temporal ordering, as we shall see.

	\noindent{\em Time crystal signatures---}We first seek for solutions in a large system using density-matrix-renormalization-group (DMRG) method. Remarkably, time crystal behaviors show up in a parameter {\em region} where the interaction strength $U$ is large enough (in units of $\hbar/t_2$) and $J/U$ is small, completely without disorder or fine-tuning. Two examples with different $U=0.5$ and $0.7$ for $J=0.2, \Delta=0.1$ are presented in Fig. \ref{fig:twochains}(b) for the system size $L=80$ on each chain. Here we consider hard-core bosons, with the initial state that one of the two legs is fully occupied, i.e. $|\psi_i\rangle = \prod_i a_i^\dagger |0\rangle$. When the ``Rabi frequency'' is perturbed by $\varepsilon=0.12$, the oscillation frequency is indeed locked to $2T$.  In supplementary materials we checked the longer time behavior for a smaller system ($L=20$) using DMRG, which shows that the amplitudes cease to decay around $t/T\in [35,45]$ and remain almost a constant. We have also checked that a slight variation of Hamiltonian parameters or the initial state does not change the $2T$ periodicity \cite{supp}. Thus, conditions (A) (B) are both met.

	\begin{figure*}[t]
		\begin{picture}(300,200)
		\put(-100,100){
		\parbox{6.7cm}{
			\boxed{
				\includegraphics[width=6.5cm]{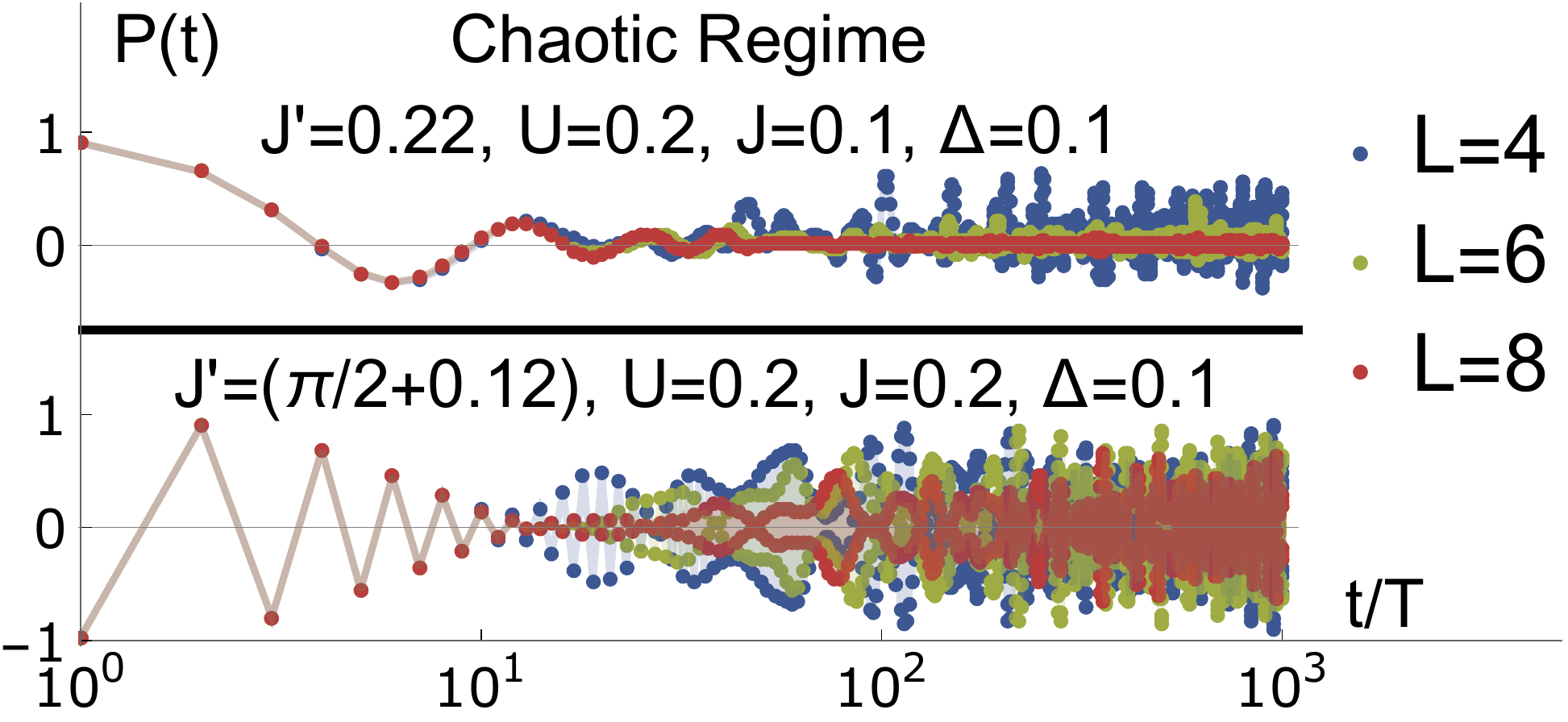} }\\
			\boxed{
				\includegraphics[width=6.5cm]{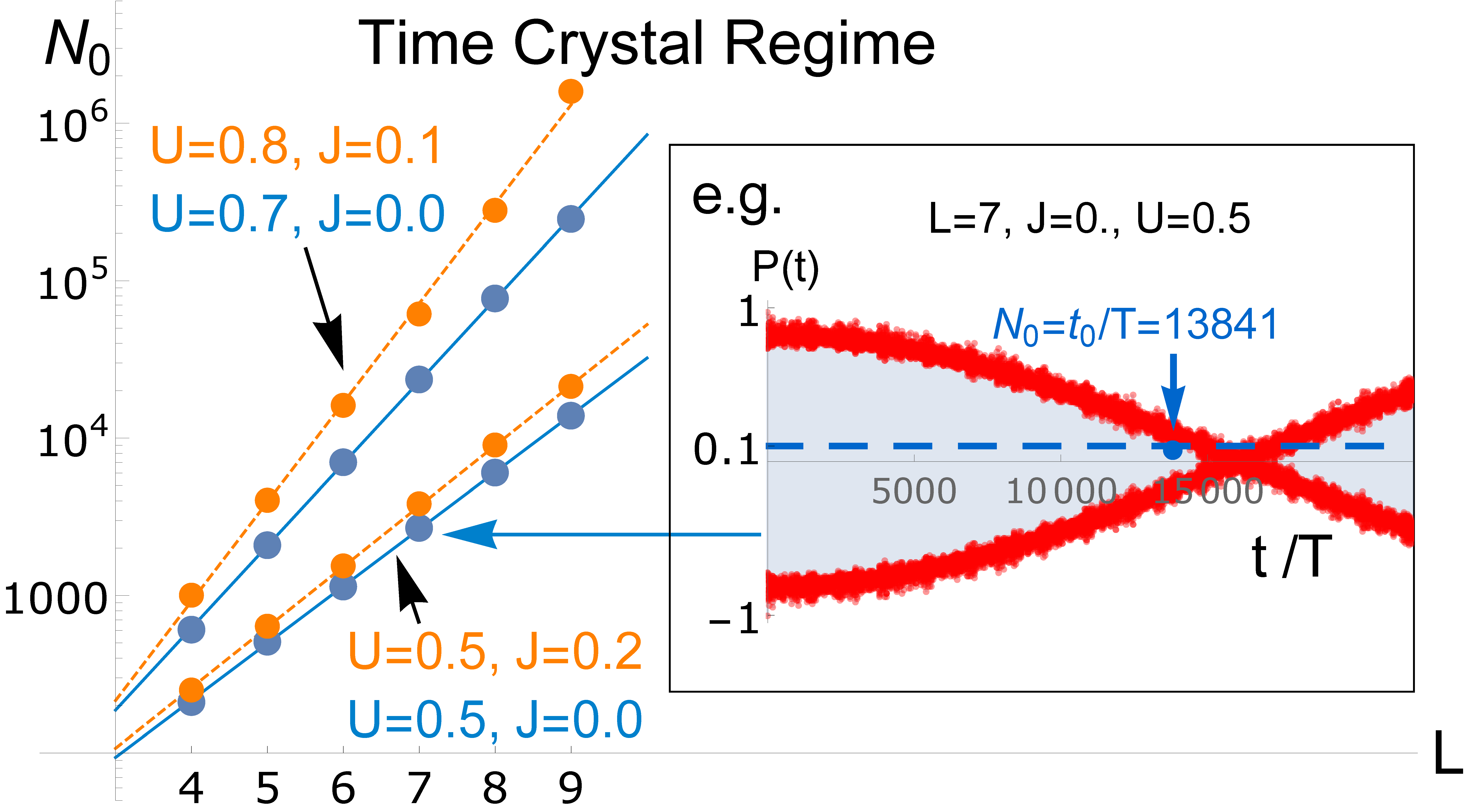}}
		}
		\parbox{3.4cm}{
			\boxed{
				\includegraphics[width=3cm]{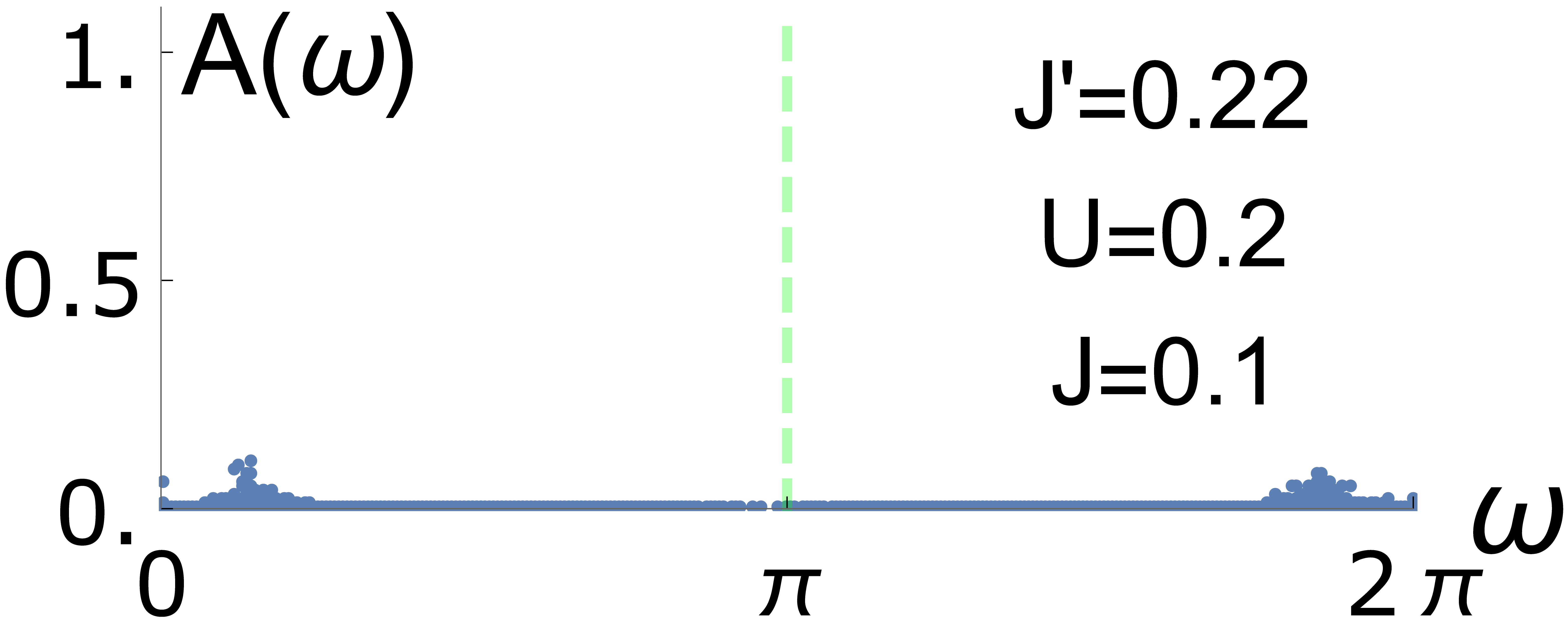}$ \, $}\\
			\boxed{
				\includegraphics[width=3cm]{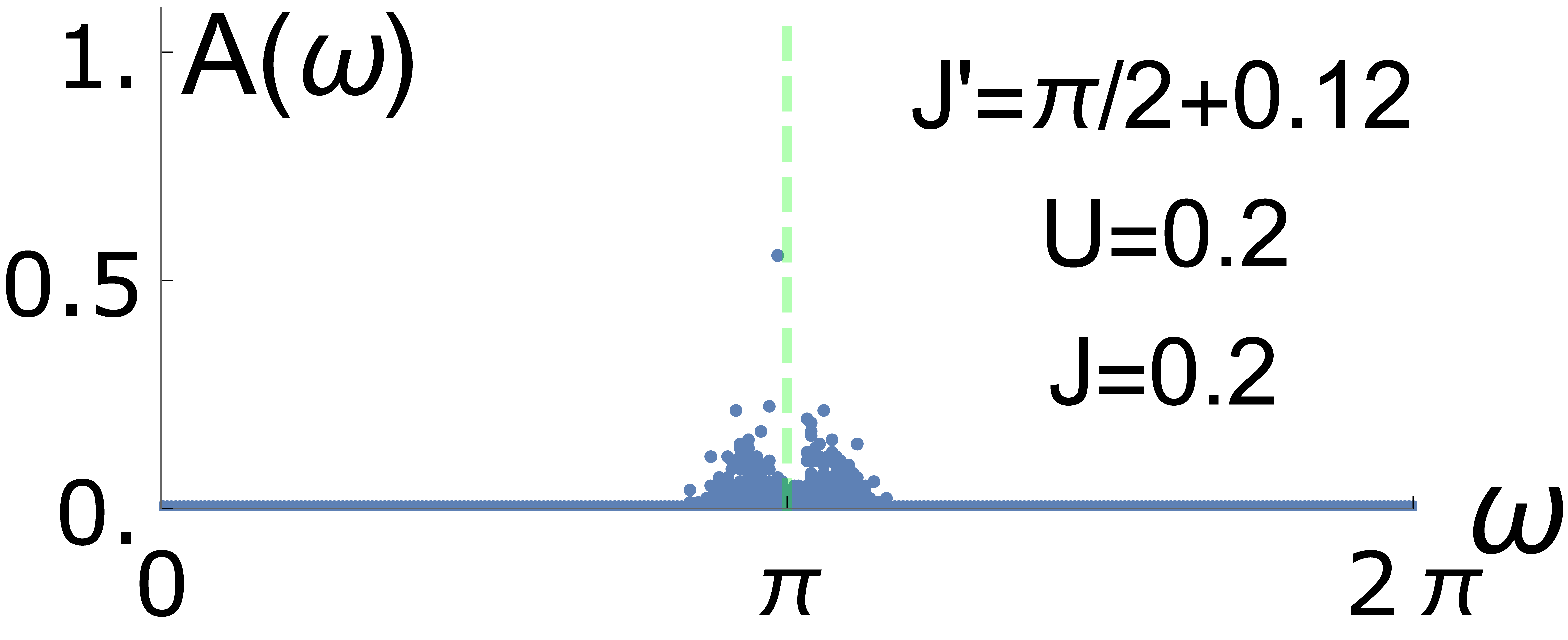}$ \, $}\\
			\boxed{
				\begin{aligned}
					\includegraphics[width=2.95cm]{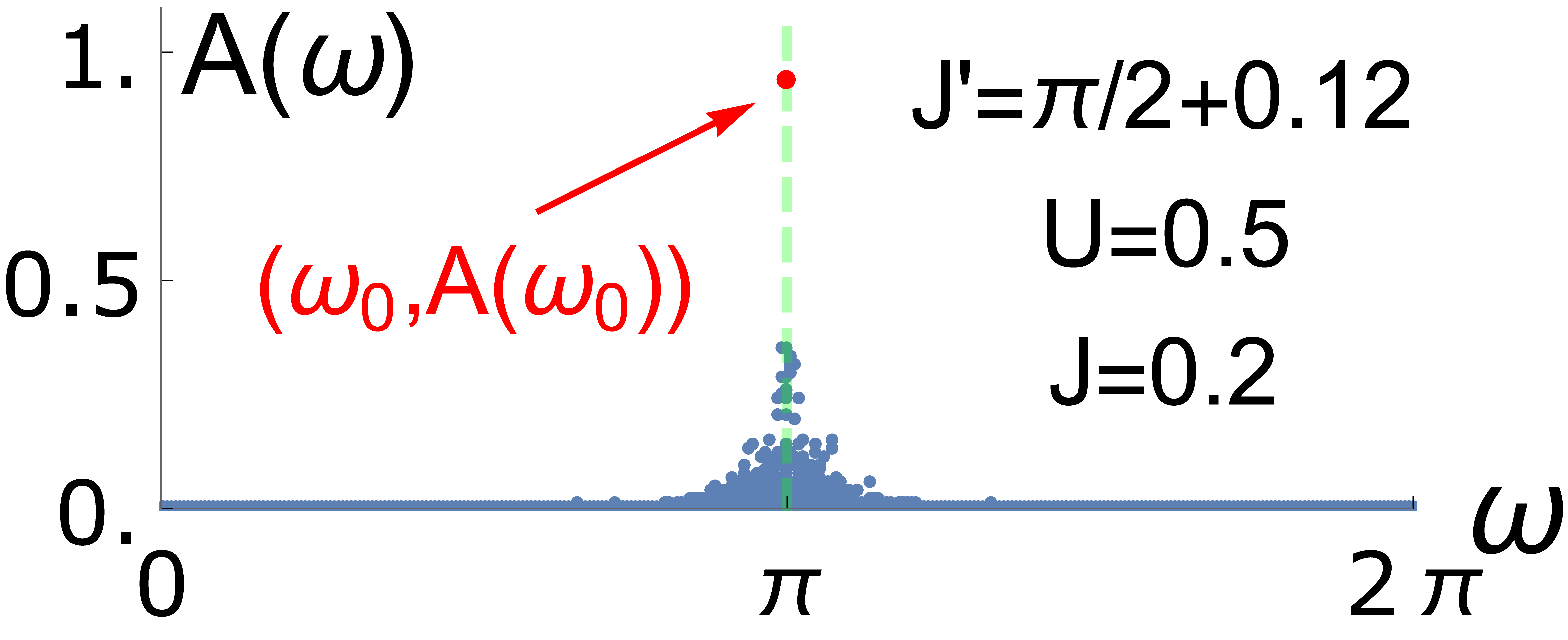}\\
					\includegraphics[width=2.95cm]{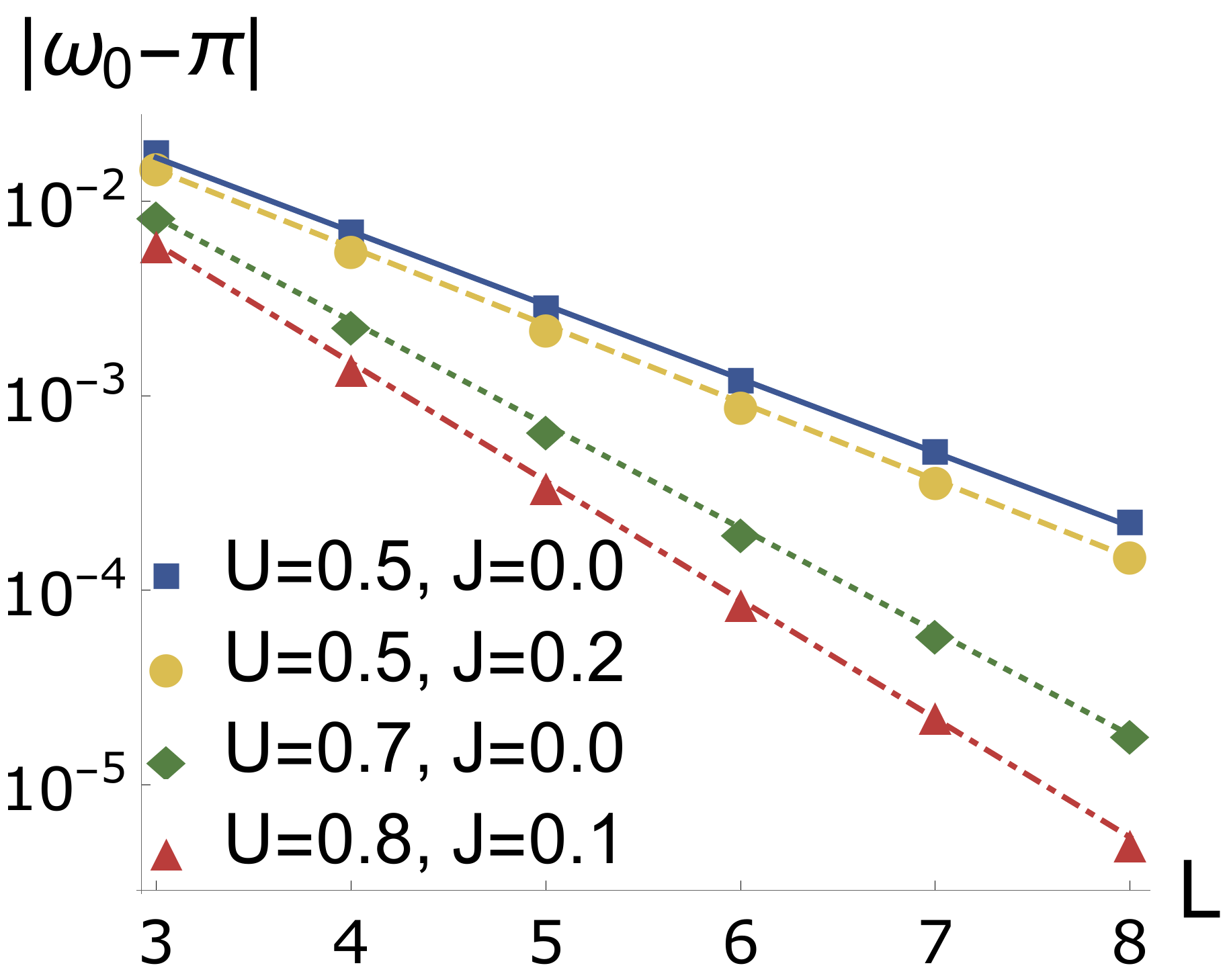}
				\end{aligned}
			$ \, $}
		}
		\parbox{3.3cm}{
			\boxed{
				\includegraphics[width=3.05cm]{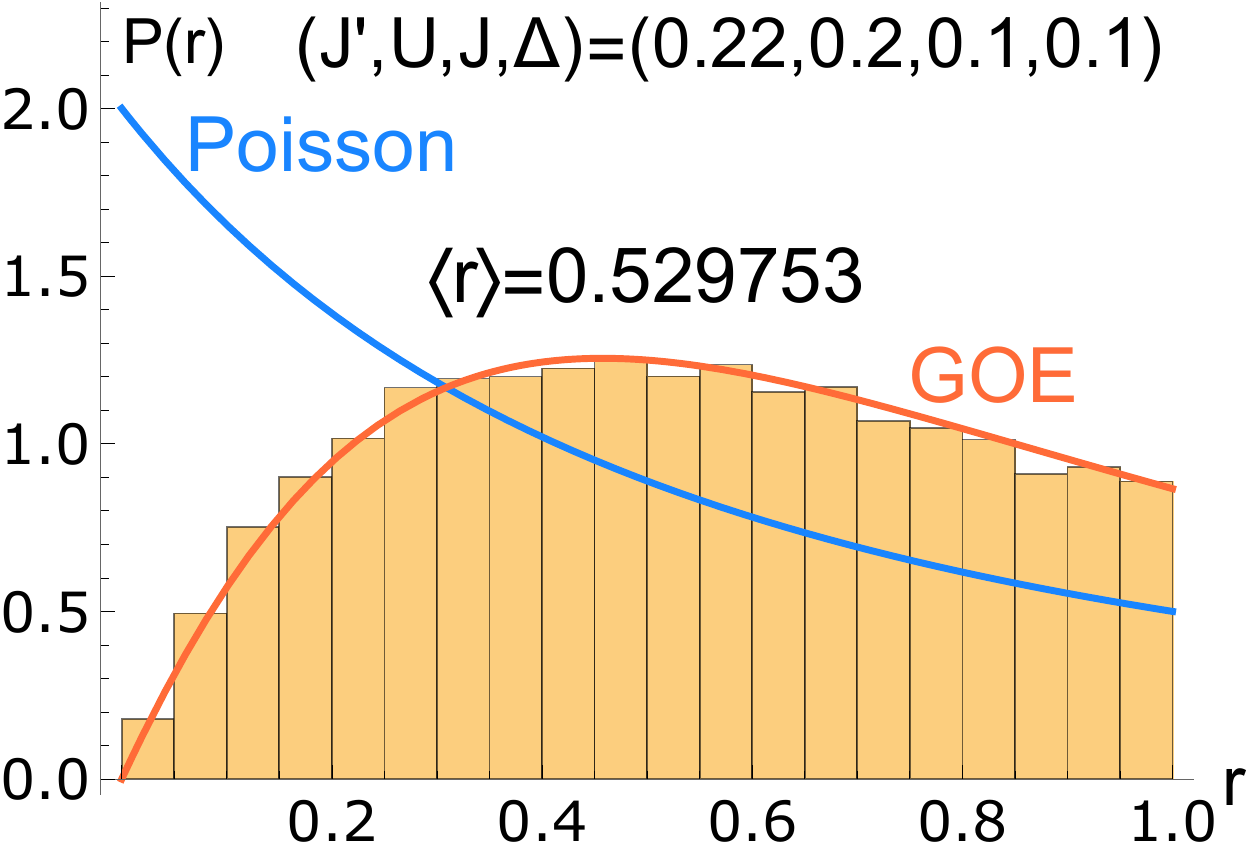} $ \, $}\\
			\boxed{
				\includegraphics[width=3.05cm]{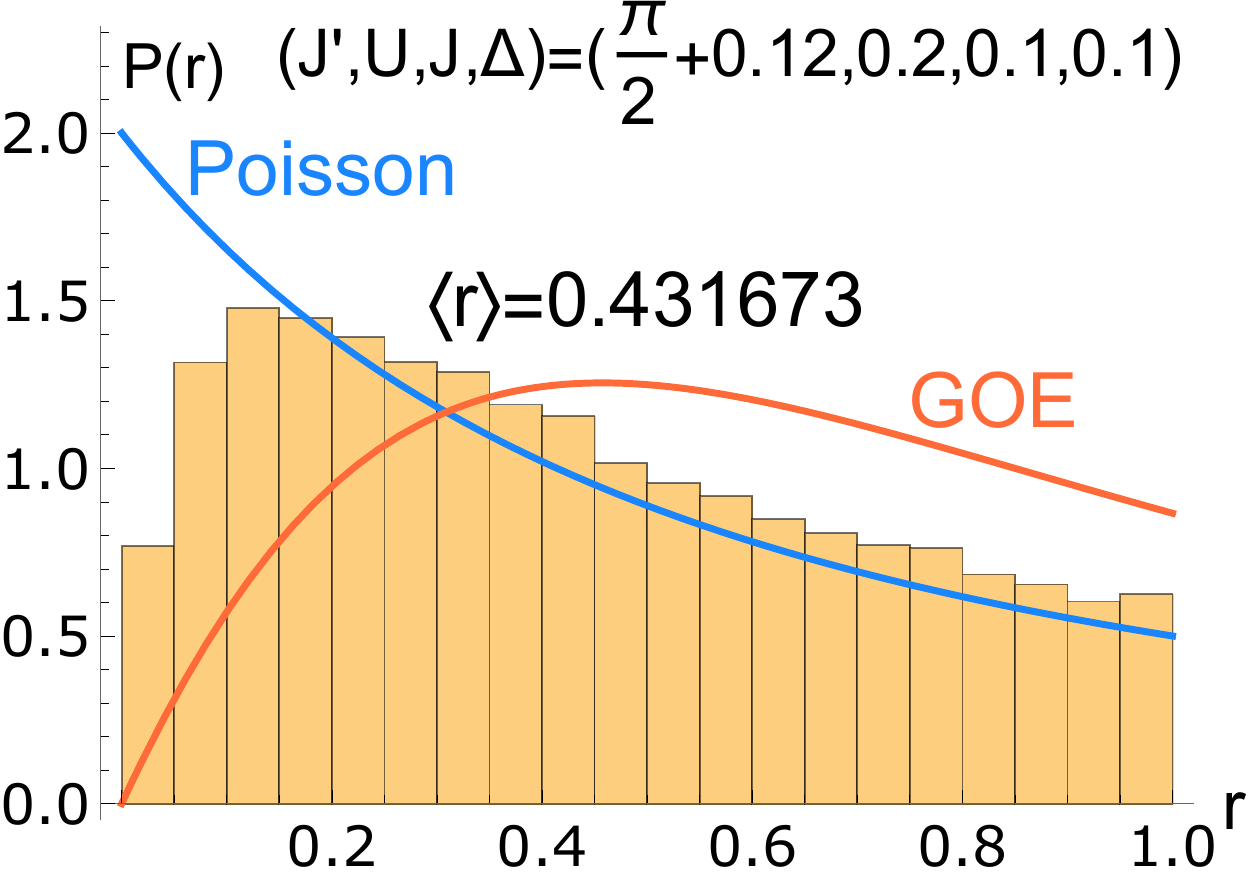} $ \, $ }\\
			\boxed{
				\includegraphics[width=3.05cm]{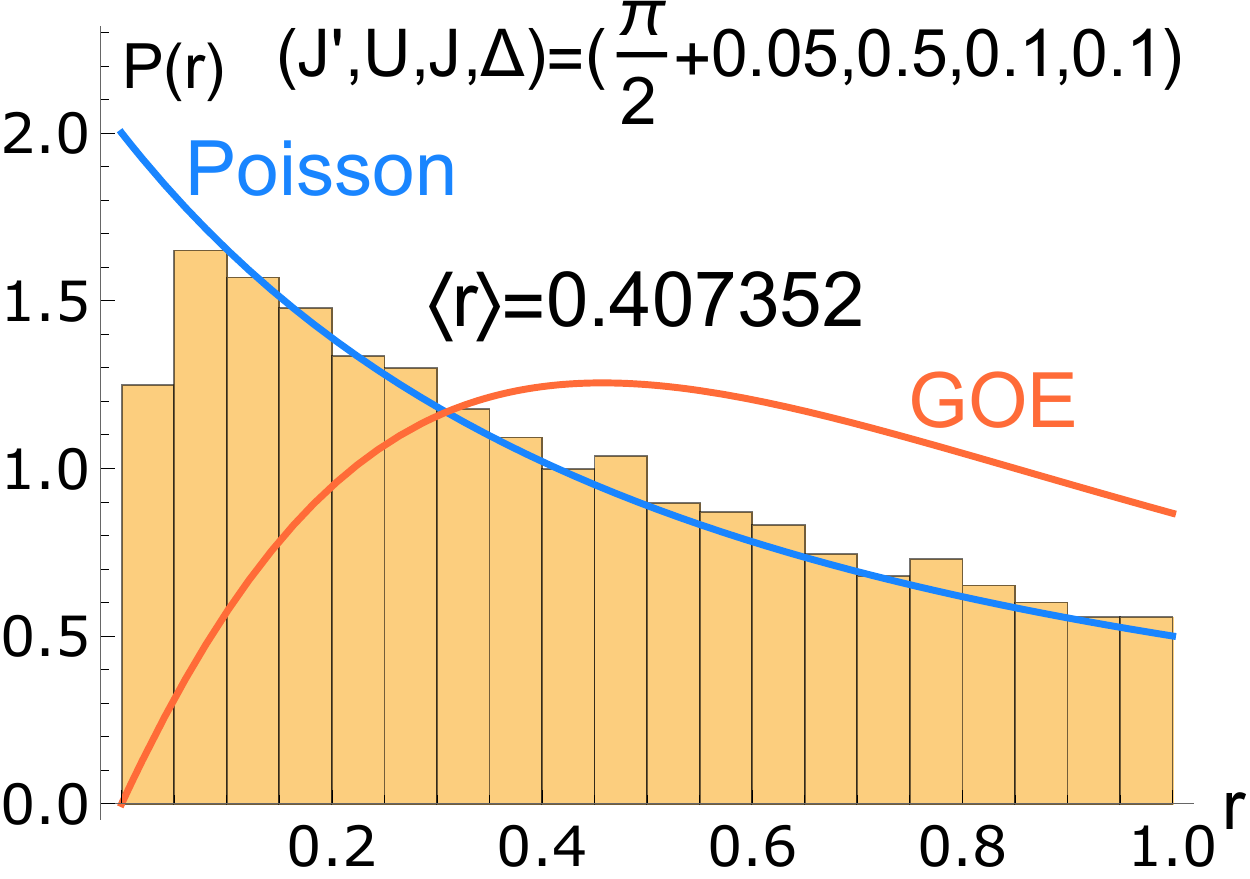} $ \, $ }
		}
		\parbox{3.75cm}{
			\boxed{
				\includegraphics[width=3.2cm]{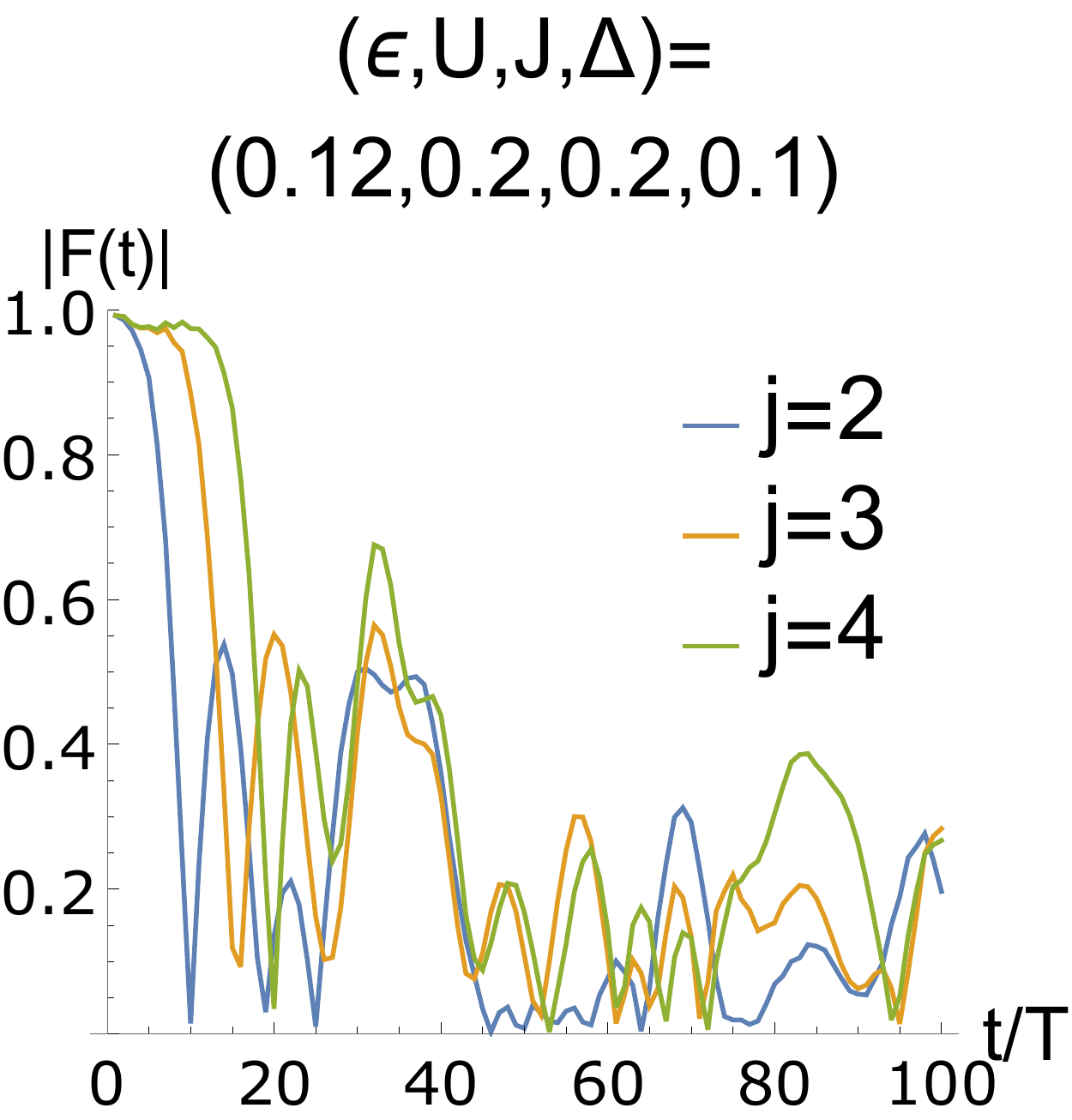}}\\
			\boxed{
				\includegraphics[width=3.2cm]{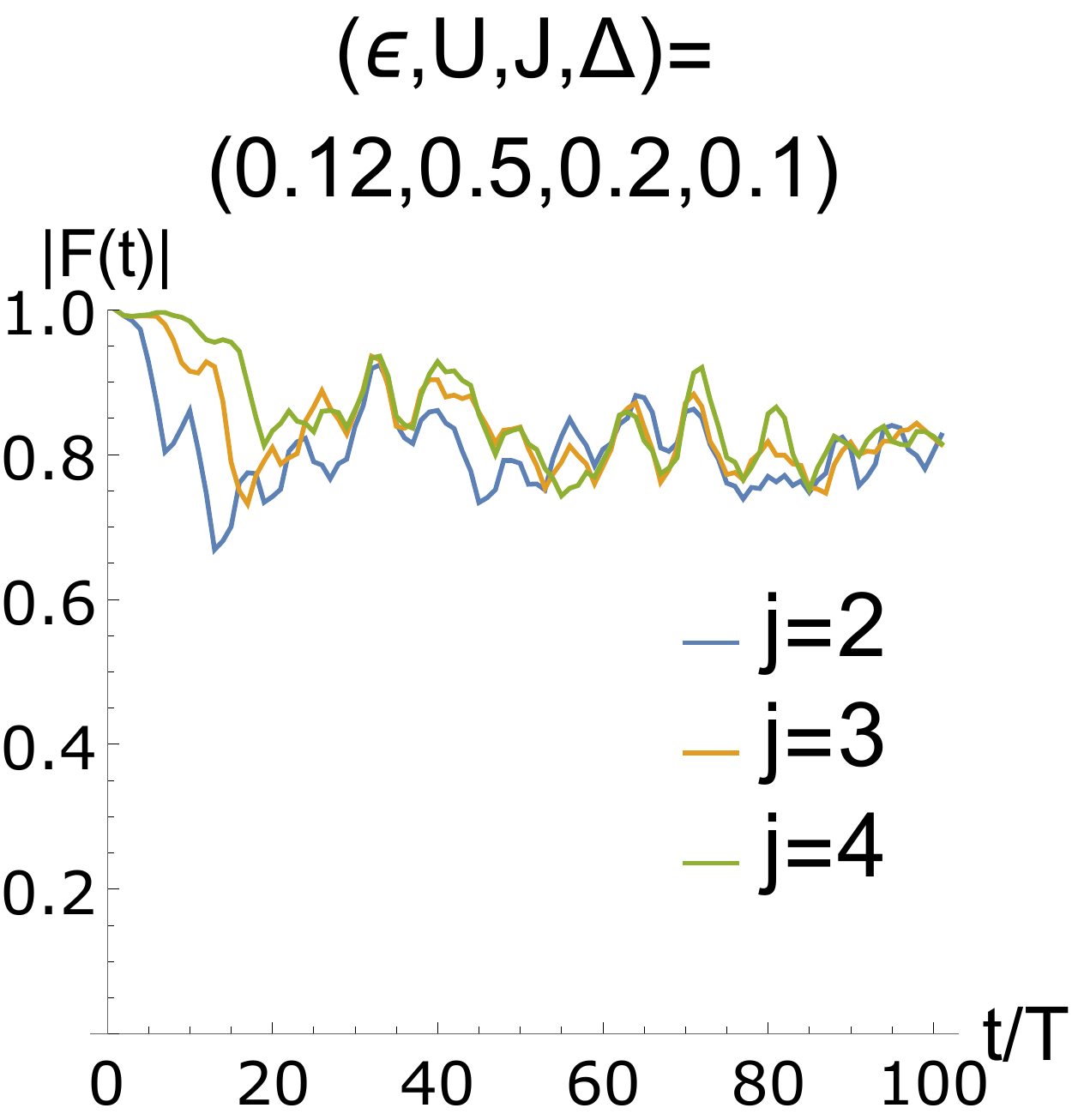}}
		}
	}
	\put(80,195){\scriptsize \boxed{a1}}
	\put(80,102){\scriptsize \boxed{a2}}
	\put(179.5,196){\scriptsize \boxed{b1}}
	\put(179.5,153.5){\scriptsize \boxed{b2}}
	\put(179.5,111){\scriptsize \boxed{b3}}
	\put(280,196){\scriptsize \boxed{c1}}
	\put(280,129){\scriptsize \boxed{c2}}
	\put(280,60){\scriptsize \boxed{c3}}
	\put(378.5,196.5){\scriptsize \boxed{d1}}
	\put(378.5,93.5){\scriptsize \boxed{d2}}
		\end{picture}
		\caption{\label{fig:tc_ntc} 
		{\bf (a1)-(a2):} Histogram of $ P(t) $. In time-crystal regime, $ P(t) $ shows an envelope modulation for the amplitude of $ 2T $-periodic oscillations. The modulation length $ N_0\equiv t_0/T $ (set by $ P(t_0) $ decreasing below $ 10\% $ of the initial value) scales exponentially with system size. 
		{\bf (b1)-(b3):} Spectral weight $ A(\omega) $ for temporal correlation functions, where $ \omega $ carries the unit $ 1/T $. (We plotted $ L=6 $ for example, and $ \Delta=0.1 $). 
		{\bf (c1)-(c3):} Distribution of level spacing ratios ($L=9$). It crosses from a GOE type deep in thermalizing regime (c1) to the Poisson limit in time crystal regime (c3).
		{\bf (d1)-(d2):} OTOC with site $ i=1 $ and for different sites $ j $'s. The system size is $ L=7 $ with periodic boundary condition. The initial state is that one of the two chains is fully occupied.
			}
	\end{figure*}

    To further understand the DMRG result and to access late time behaviors, we next turn to exact diagonalization for the same initial state with periodic boundary condition. A dramatic contrast for systems in and out of the time crystal regime is found in their finite-size scalings. Starting from isolated Rabi oscillators $ H_2=0 $, a chaotic regime is reached immediately upon turning on weak interactions $ U $, see Fig. \ref{fig:tc_ntc}(a1). After an initial period $ t/T\approx 10 $, the many-body physics sets in and the oscillation becomes non-universal for different $ L $. Especially, for weak drive $ J'=0.22 $, the oscillation amplitude decays for larger $ L $, signifying a thermalizaing behavior. However, for drivings near $ J'=\pi/2+\epsilon $, further increasing interaction strength $ U $ leads to a time-crystal regime with fixed period-$ 2T $ oscillations, consistent with DMRG results. For much later time, the oscillation amplitude shows an overall envelope shape (Fig. \ref{fig:tc_ntc}(a2) inset). But the envelope's length expands exponentially with increasing system size (see Fig. \ref{fig:tc_ntc}), indicating a constant oscillation amplitude in the thermodynamic limit and fulfilling the requirement (C).
    
    Complementary views can be provided by temporal correlation functions \cite{vedikaunpublished}, 
    \begin{eqnarray}
    C(\omega) &=& \sum_{N=-\infty}^\infty \frac{e^{-i\omega NT}}{2\pi} \sum_n \langle \omega_n|{\cal \hat{P}}(NT) {\cal \hat{P}}(0)|\omega_n\rangle \\
    &=& \sum_{mn} \delta(\omega-\omega_{mn}) A(\omega_{mn}).
    \end{eqnarray}
    Here $ {\cal \hat{P}} = \frac{1}{L}\sum_i \hat{P}_i $, $ U_F|\omega_m\rangle = e^{i\omega_m T}|\omega_m\rangle $, and the spectral weight $ A(\omega_{mn}) = |\langle \omega_m|{\cal \hat{P}}|\omega_n\rangle|^2 $, $ \omega_{mn}=\omega_m-\omega_n  $. We emphasize that a direct calculation of spectral weight $A(\omega_{mn})$ at arbitrary Floquet eigenstates gives us {\em infinite} time response characters to arbitrary initial states. The time-crystal phase is highlighted by a strong peak of $ A(\omega_0) $ at $ \omega_0 T= \pi $ (Fig. \ref{fig:tc_ntc}(b3)) corresponding to $ 2T $ periodic motions of $ P(t) $, compared with no or weak peaks in other regimes (Fig. \ref{fig:tc_ntc}(b1)-(b2)). For finite-size systems, the shrinking deviation $ |\omega_0T-\pi|\sim e^{-\alpha L}$ (Fig. \ref{fig:tc_ntc}(b3)) corresponds to the expanding modulation length $ N_0 $ for $P(t)$. 
    
    {\em Emergent Floquet-integrability---}The coupling between two chains $ H_1 $ breaks the integrability of $ H_2 $, and the linear combinations $ \alpha H_1 + \beta H_2 $ should exhibit thermalizing behaviors in late-time dynamics if localization is absent. Then, how do we understand the non-trivial dynamics in the time-crystal regime? The key point is that when the system is under strong drive, i.e. the Hamiltonian parameters are no longer much smaller than Floquet driving frequencies, the Magnus expansion of $ U_F $ is no longer dominated by the linear terms of static Hamiltonians, and it turns out that emergent Floquet integrability shows up in the time crystal regime as a property of $ U_F $.
    
    We first look at level statistics as a diagnostics of integrability~\cite{rdistribute}. Arrange the Floquet quasi-energies $\alpha_m\in (0,2\pi): U_F|\alpha_m\rangle = e^{i\alpha_m}|\alpha_m\rangle $ such that $ \alpha_{m+1}>\alpha_m $, define the level spacings $ \delta_m = \alpha_{m+1}-\alpha_m $ and further the ratios $ r_n = \max(\delta_{m},\delta_{m+1})/\min(\delta_{m},\delta_{m+1}) $, we typically end up with two distributions of $ r_n $ with probability $ P(r_n) $. In the integrable limit, such as in MBL systems, we expect a Poisson distribution $ P(r) = 2/(1+r)^2 $ with mean values $ \langle r\rangle \approx 0.386 $. Contrarily for thermalizing systems, level repulsion gives a Gaussian orthogonal ensemble (GOE) for $ P(r) = (27/4) (r+r^2)/(1+r+r^2)^{5/2} $ with the mean value $ \langle r\rangle \approx 0.536 $. From Fig. \ref{fig:tc_ntc}(c1)-(c3), we see that as one goes from thermalizing regime (c1),(c2) to deep in the time-crystal regime (c3), the distribution gradually crosses from the GOE type to the Poisson limit. 
    
    To further understand the emergent integral of motion, we compute the out-of-time-order correlators (OTOC),
    \begin{equation}\label{ft}
    F(t) = \frac{\langle W_i^\dagger (t) V_j^\dagger (0) W_i(t) V_j(0) \rangle}{\langle W_i^\dagger(t) W_i(t)\rangle \langle V^\dagger V\rangle }.
    \end{equation}
    Here $ i,j $ are site indices, and operators $ W_i, V_j $ are both chosen as local density polarization $ P_i, P_j $, for reasons specified later. The average is taken on the state of interest, i.e. the initial state. Such a correlator has the intriguing property of quantifying quantum chaos, and has been used extensively in recent works ranging from gravity theories \cite{bh} to quantum many-body systems \cite{information,protocols,mbl}. Several experimental measurements \cite{expt} have also been performed recently.
    
    For isolated Rabi oscillators with $ H_2=0 $, $ W_i(t) $ remains local and commutes with $ V_{j\ne i} $ for all time, giving a constant $ |F(t)| $. In contrast, OTOC in thermalizing systems should decay to and remain a small value \cite{mbl}. But if the system possesses integrals of motion with $ W_i, V_j $ having large overlap with them, $ F(t) $ would remain close to unity. Accordingly, we find a sharp contrast of OTOC in and out of the time-crystal phases, as shown in Fig. \ref{fig:tc_ntc}(d1) and (d2) respectively. The fact that $ |F(t)| $ for $ P_i $ remains a large value prompts us to suggest the possible form for emergent Floquet-integral of motion $I^\alpha = \sum_i k_i^\alpha \hat{P}_i$: $ \{ U_F, \hat{I}^\alpha\}_+ = O(e^{-L})\xrightarrow[]{L\rightarrow\infty} 0$, when the parameters are within time-crystal regime, where $\hat{P}_i$  is defined in Eq.~(\ref{pt}). As we do not have localizations, the configuration for the proportionality coefficients $\{k^\alpha_i\in \mathbb{C}\}$ can be extended in space. 
    
    Two caveats are in order. First, the integrals of motion in our system may not be complete, as can be reflected in the imperfect Poisson distribution in Fig.~\ref{fig:tc_ntc}(c3) and an irregular pattern of $\langle r\rangle$ when system sizes change. This resembles the ``partial thermalization'' as in mobility edge of MBL \cite{mobility1, mobility2}  or in quantum disentangled liquids \cite{qdl1,qdl2,qdl3}. Second, the characters we show differ from the typical description of ``pre-thermal time crystals'' in Ref.~\cite{nayakprethermal}, where oscillations cease to exist within fixed time regardless of system size and a longer thermalization time relies on weaker interactions. However, our time crystal phase requires strong interactions, and the temporal correlator in Fig.~\ref{fig:tc_ntc}(b3) with a dominant peak clearly dictates persisting oscillations to infinite time, as one can verify that the same histogram in the inset of Fig.~\ref{fig:tc_ntc}(a2) repeats with modulation periods $N_0$. 
	
	{\em Experimental realization and generality---}Since the time-crystal phase does not rely on the integrability of static Hamiltonians, we expect such phases to persist when the models in Eqs. (\ref{h1})-(\ref{h2}) are generalized. This is verified by the following results for experimental proposals using dipolar gases or alkaline-earth atoms with spin-SU(N) symmetry.

	Dipolar atoms \cite{Er,dy1,dy2} or molecules \cite{krb,krb2,nak,narb,rbcs} have been successfully trapped in current cold atom experiments. In our case, the interaction within each chain can be written as \cite{supp}
	\begin{equation}\label{dipolarU}
		V_{\mbox{\scriptsize dip}} = \sum_{ij} \left(U_{\mbox{\scriptsize dip}}/x_{ij}^3\right) (n_i^A n_j^A + n_i^B n_j^B)
	\end{equation}
	where $ x_{ij} $ is the distance between lattice sites $ i, j $ along a chain, and $ U_{\mbox{\scriptsize dip}} $ is the interaction strength. This term replaces the nearest-neighbor interaction proportional to $ U $ in Eq.~(\ref{h2}). In particular, one can polarize the dipolar gases along suitable directions by electric fields such that there is vanishingly small interaction between two chains \cite{supp}.

	Alternatively, using SU(N) fermions \cite{sr1,sr2,yb1,yb2}, one can engineer an ``infinite-ranged'' interaction
	\begin{equation}\label{sunU}
	V_{\mbox{\scriptsize SU(N)}} = U\sum_{m<m'} (n_m^An_{m'}^A + n_m^Bn_{m'}^B),
	\end{equation}
	where the particle at each ``site'' $ m $ interacts with all particles at other ``sites'' $ m' $. Here we have exploited the concept of ``synthetic dimensions'' where one uses the internal degree of freedom, i.e. spins $ m=-S,-S+1, \dots, S $, to play the role of different lattice sites. For atom species trapped in current experiments, the spin $ S $ can be $ 9/2 $ for $ ^{87} $Sr \cite{sr1,sr2}, or $ 5/2 $ for $ ^{131} $Yb \cite{yb1, yb2}. The SU(N) particle gains its name as the interaction (\ref{sunU}) among different spin species preserve the SU(N) symmetry. One therefore only needs a tight double-well potential accommodating totally $ N=(2S+1) $ particles in its lowest orbital state if we have half-filling in the initial state.
	
\begin{figure}
		[h]
		\parbox{4cm}{\includegraphics[width=3.8cm]{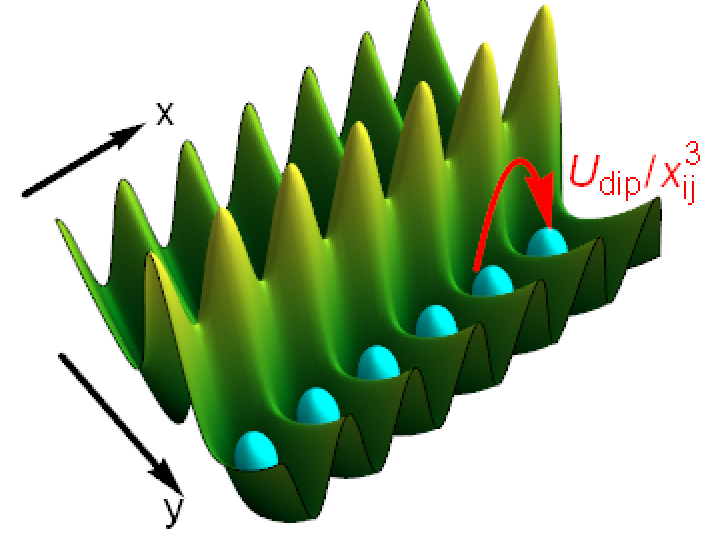}}
		\quad
		\parbox{4cm}{\includegraphics[width=3.8cm]{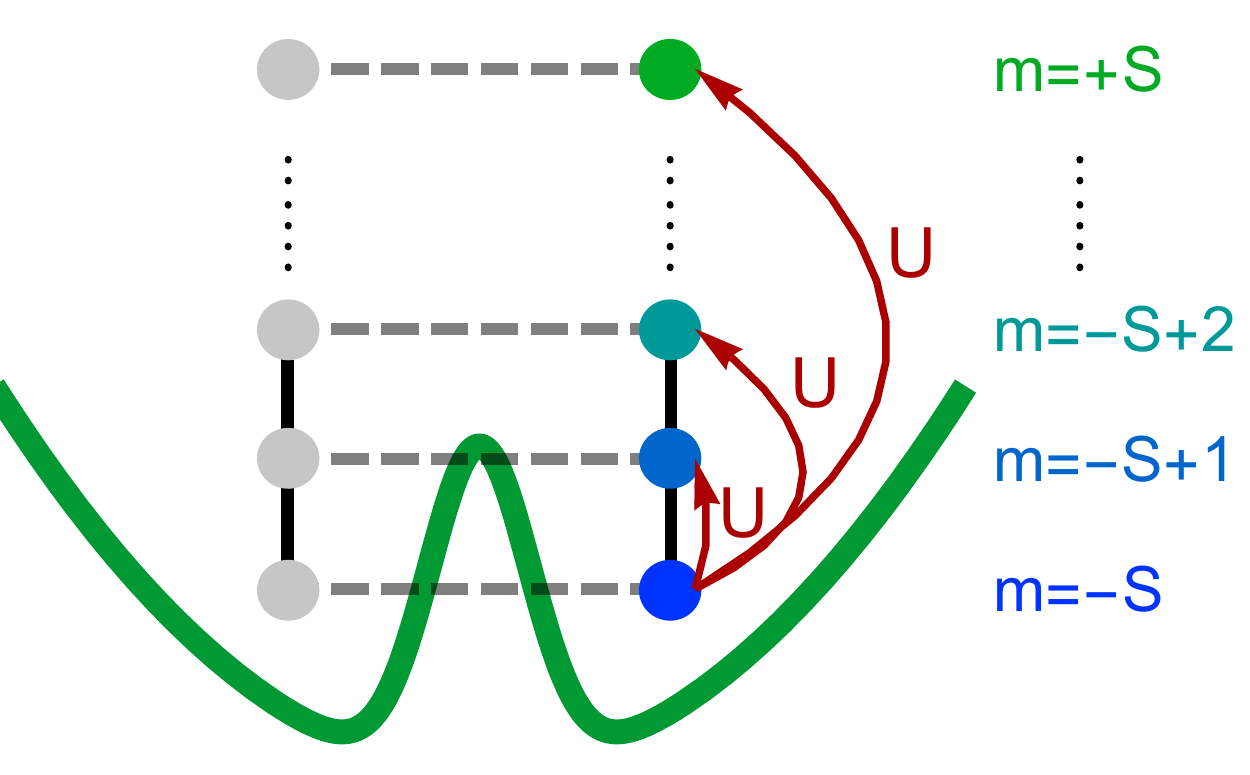}}
		\\
		\parbox{4cm}{\includegraphics[width=3.8cm]{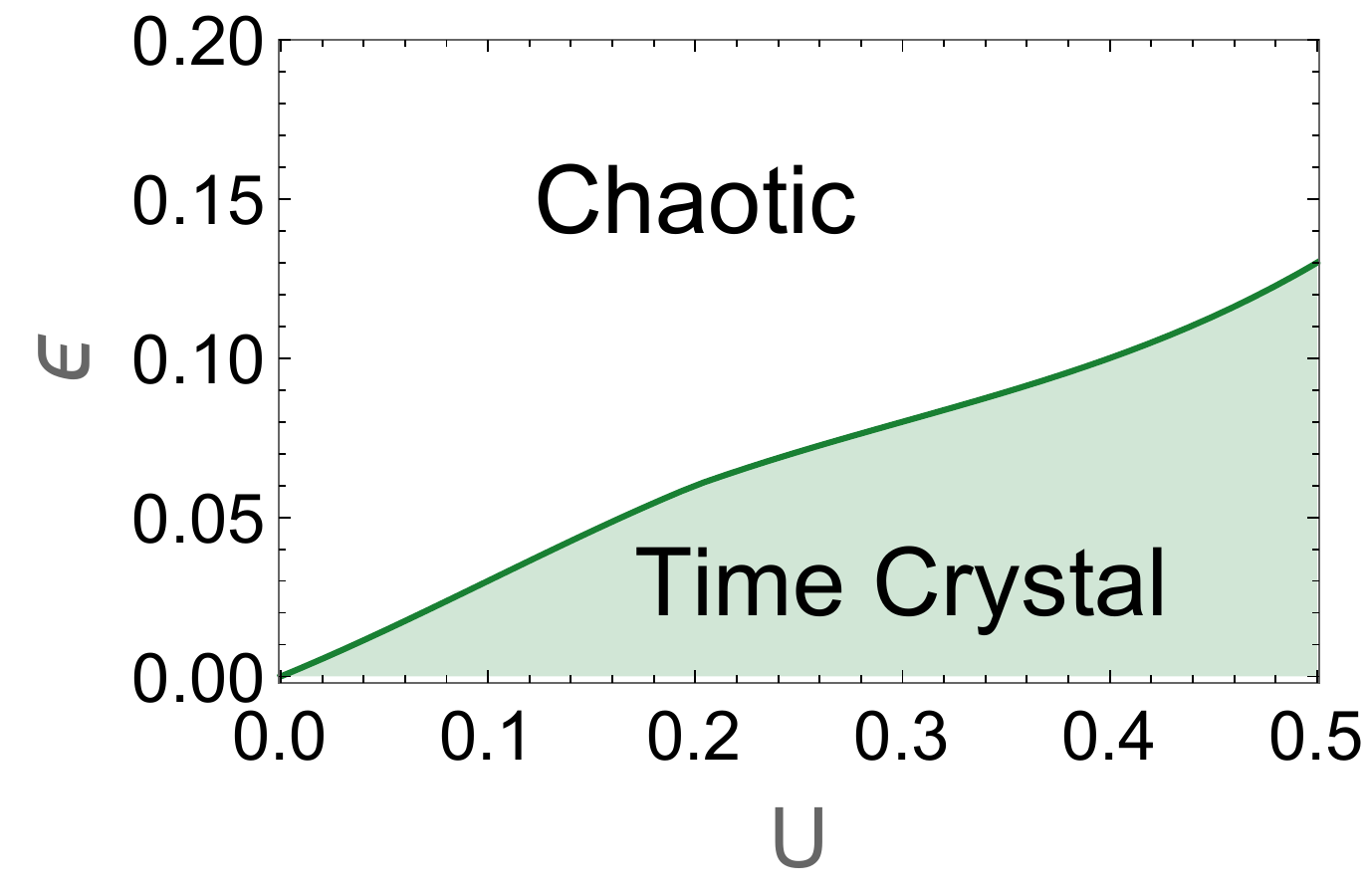}} \quad
		\parbox{4cm}{\includegraphics[width=3.8cm]{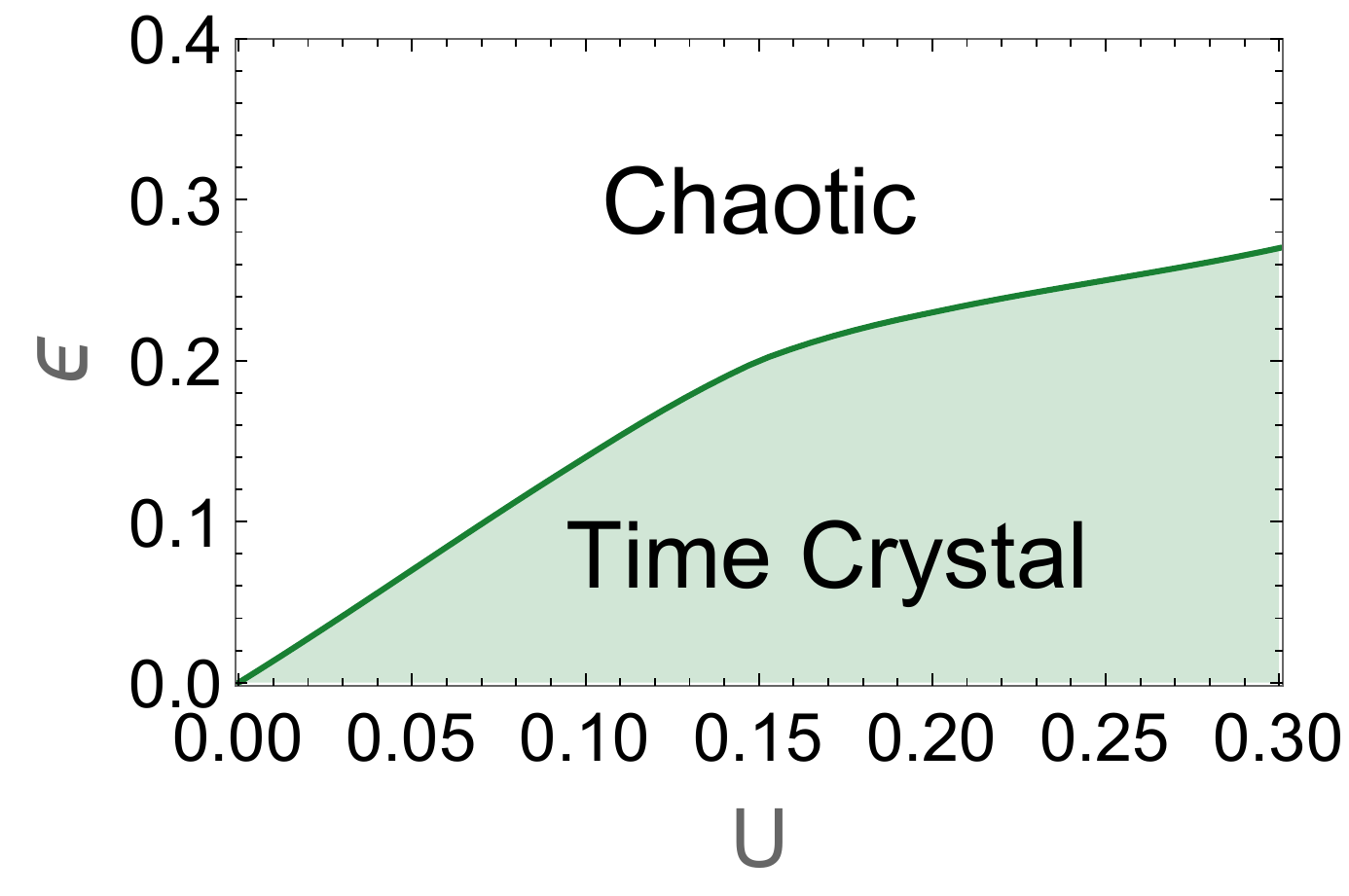}}
		(a)\qquad\qquad\qquad\qquad\qquad\qquad (b)
		\caption{\label{fig:experiment}
			The experimental set-ups and phase diagrams for fermionic (a) dipolar gases ($ J=0.4U, L=6 $) and (b) SU(N) particles ($ J=0, L=10 $ for $ ^{173} $Yb) with open boundary conditions. The phase boundary is set to that the ``envelope'' height of the oscillation, as shown in the inset of the Fig. \ref{fig:tc_ntc}(a), remains above (or below) 50\% for the time crystal (or chaotic) phase during the first 200 periods. 
			}
	\end{figure}

	We refer the readers to Supplemental Material for details regarding lattice set-up, quench process, and parameter estimations. Here we present a phase diagram for each of these two cases in Fig.  \ref{fig:experiment}(a) and (b) respectively. We clearly see that time crystal phases are stabilized by strong interactions.

	\noindent{\em Conclusion ---} We have shown through explicit models that a stable time crystal phase exists without the need for fine tuning or localization by disorder. The exponential scaling of the modulation length with respect to system size, together with the dependence on {\em strong} interaction strength, imply that the clean-Floquet time crystal phase is different from the usual pre-thermal state \cite{nayakprethermal}. The existence of such a phase is of genuine dynamical origin, where certain integrals of motion emerge in the Floquet operator instead of being in the static Hamiltonian. Therefore, it points to a tantalizing possibility of using dynamical process  to preserve quantum information. Finally, as being confirmed in the experimental proposals, the time-crystal behavior is not restricted to a specific model. Thus, it is intriguing to generalize the present discussions to systems with more complexity in parallel to usual spatial crystals. Studying time crystals in various clean systems will surely yield new principles and phenomena of non-equilibrium nature.

	\noindent{\em Acknowledgement---}The authors wish to thank Vedika Khemani for introducing to us the spectral function diagnostic method, and V. Khemani, Norman Yao, Dominic Else, Xiaopeng Li and Soonwon Choi for comments and suggestions on thermalization issues. We also would like to thank Shivaji Sondhi, Yi-Zhuang You, Meng Cheng and Joel Moore for discussions. This work is supported by U.S. ARO (W911NF-11-1-0230) and AFOSR (FA9550-16-1-0006) (B.H. and W.V.L.), and Overseas Collaboration Program of NSF of China (No. 11429402) sponsored by Peking University (W. V. L.), and the DFG within the Cluster of Excellence NIM (Y.-H. W.).

	\pagebreak
	
	\widetext
	\clearpage
	
	\setcounter{equation}{0}
	
	\begin{center}
		{\bf\Large Supplemental Material}
	\end{center}

		\section{Mapping to spin models}
		In this section, we rewrite the Hamiltonians in Eqs.~(2), (3) in the main text as spin models. It is well known that a Hubbard-type of interacting chain can be mapped into a spin-$ 1/2 $ XXZ spin chain \cite{xiaohuangshu}. For hard-core bosons
		\begin{equation}
		[a_i, a_j^\dagger] = (1-2n_i)\delta_{ij}
		\end{equation}
		the mapping is
		\begin{equation}
		a_i^\dagger = S_i^x + i S_i^y, \qquad 
		n_i = a_i^\dagger a_i = S_i^z + 1/2.
		\end{equation}
		Then the Hamiltonians become
		\begin{eqnarray}
		t_1: \quad & \quad H_1 = 2J' \sum\limits_{i=1}^L (S_{Ai}^x S_{Bi}^x + S_{Ai}^y S_{Bi}^y)\\
		t_2: \quad & \quad H_2 = \sum\limits_{i=1}^L \sum\limits_{\mu=A,B} \left[ 2J (S_{\mu i}^x S_{\mu i+1}^x + S_{\mu i}^y S_{\mu i+1}^y)+ U S_{\mu i}^z S_{\mu i+1}^z 
		\right]
		+ \sum\limits_{i=1}^L \left[(U+\Delta)S_{Ai}^z + (U-\Delta)S_{Bi}^z \right]
		\end{eqnarray}
		On the other hand, for fermions
		\begin{equation}
		\{ f_i, f_j^\dagger \} = \delta_{ij},
		\end{equation}
		we can perform a Jordan-Wigner transform and map to the spin operators,
		\begin{equation}
		S_i^x = (f_i + f_i^\dagger) e^{i\pi\sum_{j=1}^{i-1}n_j},\quad
		S_i^y = i(f_i-f_i^\dagger) e^{i\pi\sum_{j=1}^{i-1}n_j},\quad
		S_i^z = n_i - 1/2.
		\end{equation}
		Then we have a non-local XXZ type of model due to the inter-chain coupling during $ t_1 $,
		\begin{eqnarray}
		t_1 \quad & \quad H_1 = 2J'\sum\limits_{i=1}^L (S_{Ai}^x S_{Bi}^x + S_{Ai}^y S_{Bi}^y) e^{i\pi\sum_{j=Ai,\dots,AL, B1}^{Bi-1} (S_i^z + 1/2)}\\
		t_2 \quad & \quad H_2 = \sum\limits_{i=1}^L \sum\limits_{\mu=A,B} \left[ 2J (S_{\mu i}^x S_{\mu i+1}^x + S_{\mu i}^y S_{\mu i+1}^y)+ U S_{\mu i}^z S_{\mu i+1}^z 
		\right]
		+ \sum\limits_{i=1}^L \left[(U+\Delta)S_{Ai}^z + (U-\Delta)S_{Bi}^z \right].
		\end{eqnarray}
		Here $ H_2 $ is fully identical to the situation for hard-core boson. The non-local phase in $ H_1 $ reflects the fact that when we interpret the inter-chain coupling in terms of one-dimensional interactions in the chain $ (A1, A2, \dots, AL, B1,\dots,BL) $, it is a non-local one.
		
		The situation is simplified when the tunneling within each chain is suppressed, $ J=0 $ in Eq.~(3) in the main text, and with the onsite constraint $ n_i^A+n_i^B = 1 $. Then, for both hard-core bosons and fermions we can regard the two chains as two spin states, and perform the mapping
		\begin{equation}
		S_i^x = \frac{1}{2}(b_i^\dagger a_i + a_i^\dagger b_i),\quad
		S_i^y = \frac{i}{2}(b_i^\dagger a_i - a_i^\dagger b_i),\quad
		S_i^z = \frac{1}{2}(n_i^A - n_i^B), \quad
		1=n_i^A + n_i^B,
		\end{equation}
		where the last one is a constraint. Here $ a_i, b_i $ can be either bosons or fermions. Then the model becomes
		\begin{eqnarray}
		t_1 \quad & \quad H_1 = 2J'\sum\limits_{i=1}^L S_i^x,\\
		t_2 \quad & \quad H_2 = 2U\sum\limits_{i=1}^L \left( S_i^z S_{i+1}^z + \frac{1}{4}\right) + \Delta \sum\limits_{i=1}^L S_i^z
		\end{eqnarray}
		
		\section{Details and more results of DMRG calculations}
		
		\begin{figure}[h]
			\includegraphics[width=0.7\textwidth]{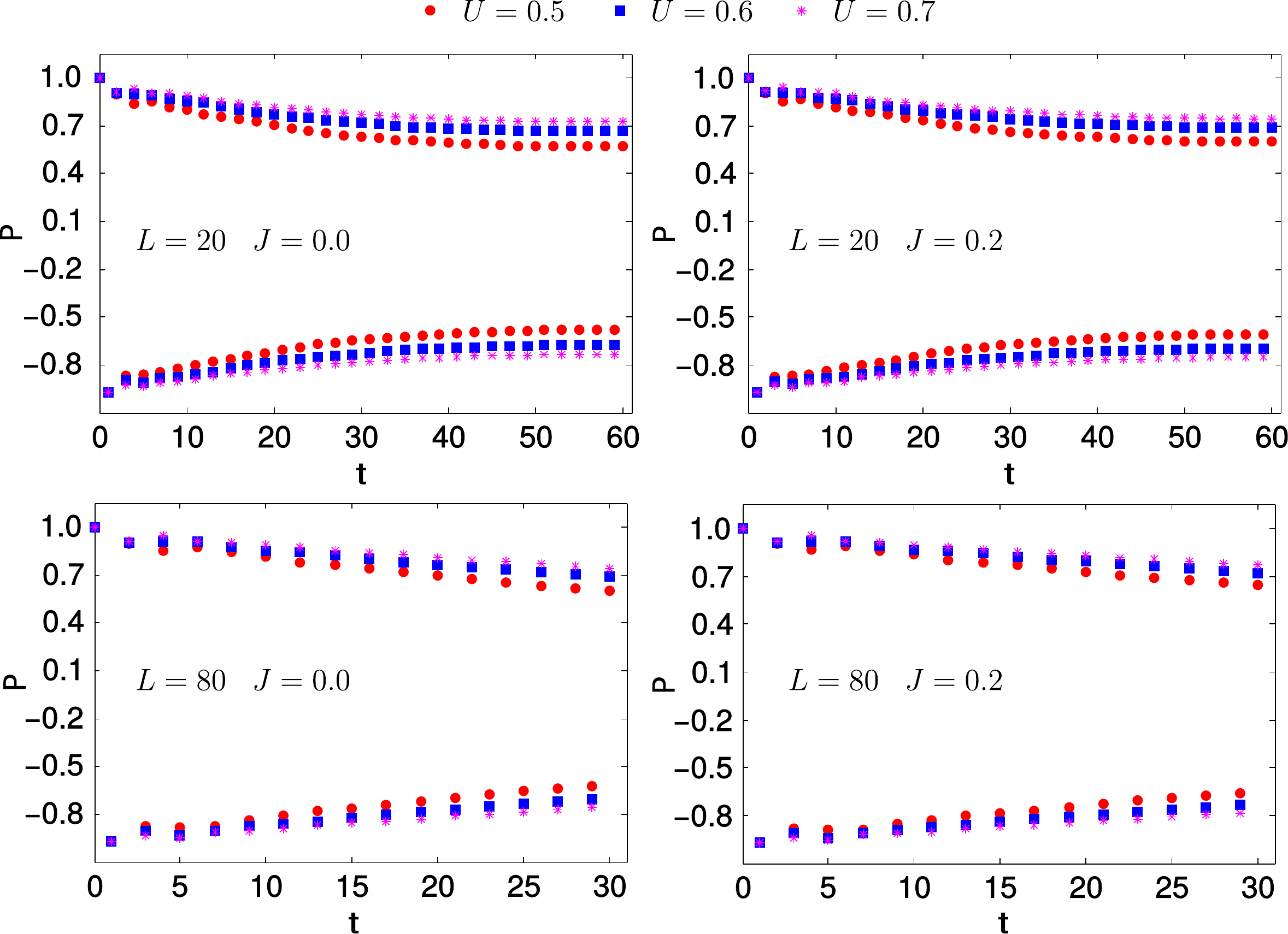}
			\caption{DMRG results of the density polarization. $J'=\pi/2+0.12,\Delta=0.1$ in all cases and $J,U$ as given as legends. The initial state is given by Eq. \ref{stateinitial} with $\alpha=0$. The accuracy for data points are over $95\%$. By studying the evolution of the smaller system $L=20$ for a long time, we see clearly that the initial decay gradually stops after $40{\sim}50$ periods.}
			\label{FigureS1}
		\end{figure}

		\begin{figure}[h]
			\includegraphics[width=0.7\textwidth]{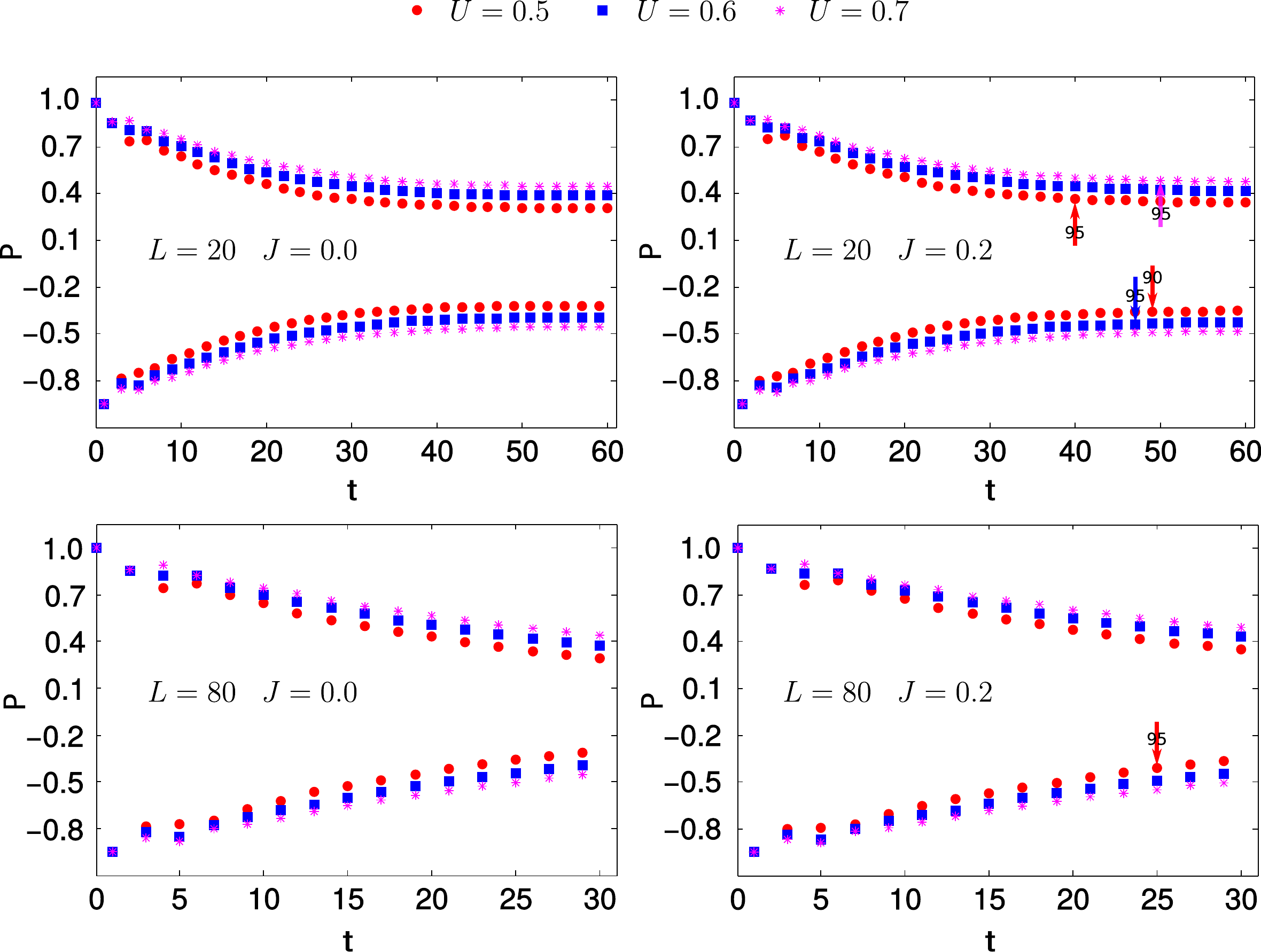}
			\caption{DMRG results for the density polarization. $J'=\pi/2+0.12,\Delta=0.1$ in all cases and $J,U$ as given as legends. The initial state is given by Eq. \ref{stateinitial} with $\alpha=0.1$ (for comparison with Fig. \ref{FigureS1}). The accuracy for most data points are over $95\%$. Similar to the results in Fig. \ref{FigureS1}, we also find that the initial decay gradually stops after $40{\sim}50$ periods.}
			\label{FigureS2}
		\end{figure}
		
		\begin{figure}
			\includegraphics[width=0.7\textwidth]{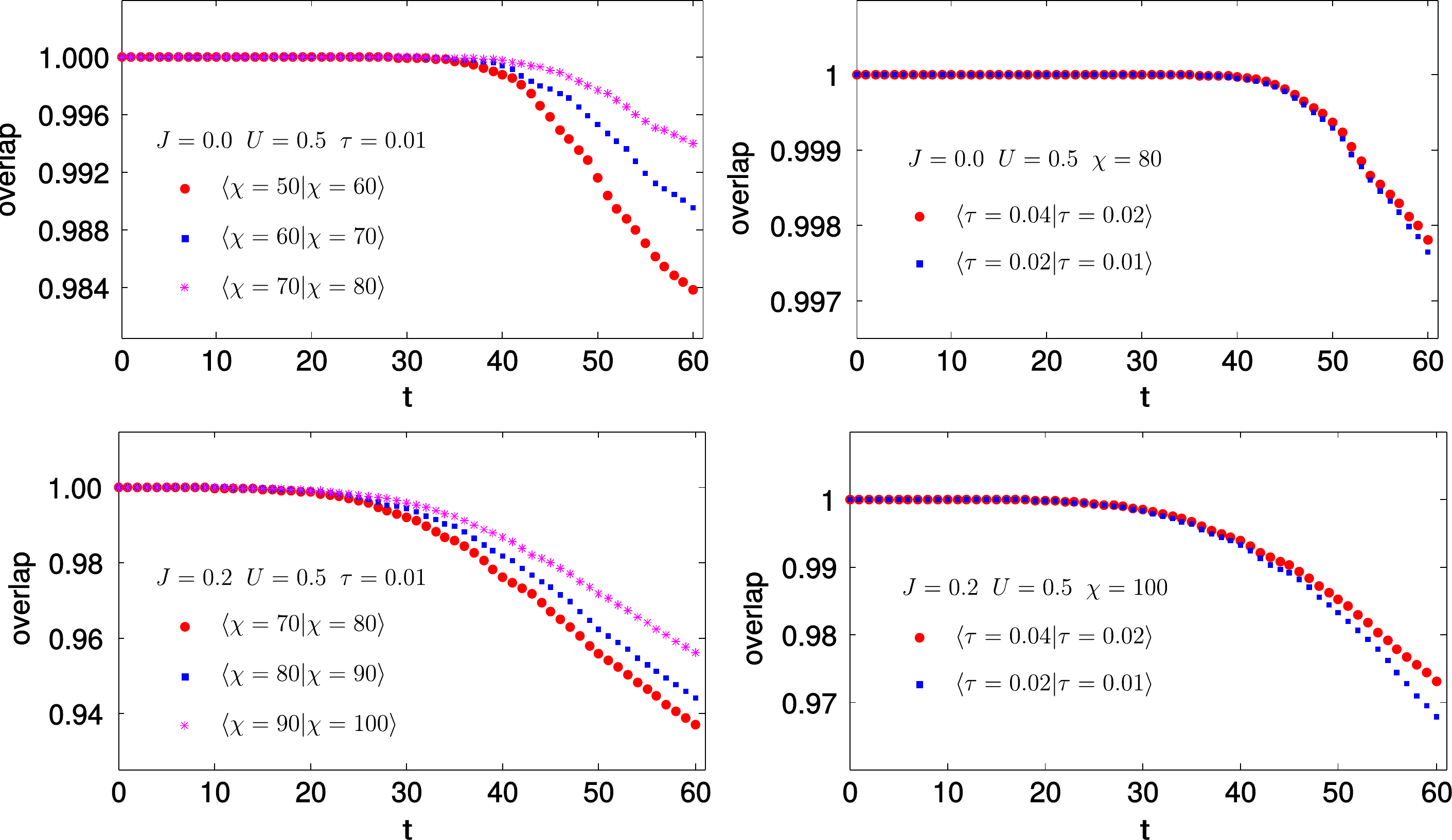}
			\caption{DMRG results of the density polarization in the $L=20$ system. $J'=\pi/2+0.12,\Delta=0.1$ in all cases and $J,U$ as given as legends. The initial state is given by Eq. \ref{stateinitial} with $\alpha=0$. The accuracy is quantified by the overlaps between time-evolved states with different bond dimension $\chi$ and time step $\tau$. The left two panels show that the overlaps approach $1$ as the bond dimension $\chi$ increases. The right two panels show that $\tau=0.04,0.02,0.01$ give similar results so $\tau=0.01$ is sufficient for our calculations.}
			\label{FigureS3}
		\end{figure}

		The calculation burdens for exact diagonalization (ED) and density-matrix-renormalization-group \cite{White1992,Schollwock2011} have quite opposite characters regarding system sizes and the number of periods. For ED, the size of Hilbert space grows exponentially with the length of the chain. But once the exact eigenstates and eigenvalues are obtained, the evolution operator at any moment can be immediately obtained. In contrast, for DMRG methods, the calculation time grows linearly for larger system sizes when the ``bond dimension" (to be discussed below) is fixed. The restriction is that the bond dimension required for achieving sufficient accuracy grows rapidly with the number of periods to be simulated. In sum, ED is suitable for checking late-time dynamics for small systems, while DMRG can give the early time behaviors for large systems. We next discuss the details and more numerical results of DMRG analysis.
		
		For a system with $L$ sites, we denote the local basis states as $|\sigma_{i}\rangle$ ($i\in[1,2,\cdots,L]$) so the many-body basis states are $|\sigma_{1},\sigma_{2},\cdots,\sigma_{L}\rangle$. The DMRG method is based on representing a physical state $|\Psi\rangle$ in the matrix product state (MPS) form as
		\begin{eqnarray}
		|\Psi\rangle = \sum_{\{s_{i}\}} M^{\sigma_{1}}_{1} M^{\sigma_{2}}_{2} \cdots M^{\sigma_{L}}_{L} |\sigma_{1},\sigma_{2},\cdots,\sigma_{L}\rangle
		\end{eqnarray}
		where $M^{\sigma_{i}}_{i}$ are matrices and the products $M^{\sigma_{1}}_{1} M^{\sigma_{2}}_{2} \cdots M^{\sigma_{L}}_{L}$ are scalars. The maximal dimension of these matrices is called the MPS bond dimension $\chi$. An arbitrary state can be written as MPS if one takes $\chi\rightarrow\infty$.
		
		The DMRG time evolution algorithm can be applied most simply if the Hamiltonian of a system contains only onsite and nearest neighbor terms. To this end, we interpret our system as a one-dimensional chain with four states on each site: no boson, one boson in state $A$, one boson in state $B$, one boson in each of the states $A$ and $B$. The initial state is chosen as a product state of sites $ i $
		\begin{equation}
		\label{stateinitial}
		|\Psi_{\rm ini}\rangle = \prod_{i}|\phi_{i}\rangle, \qquad |\phi_{i}\rangle=\cos(\alpha)|A_{i}\rangle+\sin(\alpha)|B_{i}\rangle
		\end{equation}
		with the same angle $\alpha$ used for all sites. This kind of states can be represented easily as MPS. The time evolution operator $\exp(-iHt)$ for an interval $t$ is divided to $M$ steps as $\prod^{M}_{k=1}\exp(-iH\tau)$ with $\tau=t/M$. The Hamiltonian contains only onsite and nearest neighbor terms so it can be written as $H=H_{e}+H_{o}$ where the two terms act on the even and odd lattice sites respectively. The one-step operator $\exp(-iH\tau)$ is split using a fifth order Trotter-Suzuki decomposition as
		\begin{eqnarray}
		\exp(-iH\tau) &=& \exp(-iH_{o}\theta\tau/2) \exp(-iH_{e}\theta\tau) \exp[-iH_{o}(1-\theta)\tau/2] \nonumber \\
		&\times& \exp[-iH_{e}(1-2\theta)\tau] \exp[-iH_{o}(1-\theta)\tau/2] \exp(-iH_{e}\theta\tau) \exp(-iH_{o}\theta\tau/2) + {\mathcal O}(\tau^5)
		\end{eqnarray}
		with $\theta=(2-2^{1/3})^{-1}$. The terms on the right-hand side can be obtained following standard procedure and written as matrix product operators (MPOs). When an MPO is multiplied on an MPS, the result is still an MPS but with an increased bond dimension. 
		
		The non-commutativity of $H_{e}$ and $H_{o}$ is the first approximation in the DMRG time evolution algorithm. For a fixed time step $\tau$, this error accumulates as the number of steps $M$ grows so we can not simulate for a very long time. As the application of an MPO on an MPS increases the bond dimension $\chi$, we need to truncate the result to make sure that $\chi$ stays below a reasonable value, which is the second source of error in the calculations. To make sure that the time-evolved states are sufficiently accurate, we generally perform multiple calculations using different $\chi$ and $\tau$. 
		
		Figs. \ref{FigureS1} and \ref{FigureS2} present our results of the $L=20,80$ systems. We choose the parameters $J'=\pi/2+0.12,\Delta=0.1$ in all cases and study multiple combinations of $J,U$ as indicated in the figures. The initial state is set as $\alpha=0$ in Fig. \ref{FigureS1} and $\alpha=0.1$ in Fig. \ref{FigureS2}. For each case, we only show the evolution of $P$ for the largest $\chi$ that has been used. To gauge the reliability of these results, we define accuracy at a particular time as the overlap between the time-evolved states obtained using the largest and the second largest $\chi$'s. An arrow attached with $95$ ($90$) indicates the {\em last} position where the overlap is larger than $95\%$ ($90\%$). If there is no arrow with $95$ ($90$) for a certain $U$ in a panel, all the overlaps in this panel for this $U$ are larger than $95\%$ ($90\%$). For all the cases in Figs. \ref{FigureS1} and \ref{FigureS2}, the density polarization $P$ decays at the beginning. For the $L=20$ system, we can simulate up to $60$ periods and the results demonstrate that the initial decay gradually stops after $40{\sim}50$ periods. For the $L=80$ system, we have only studied $30$ perioids due to the higher computational cost.
		
		To give a more detailed analysis of the accuracy, we show the results for $L=20$ at different $\chi$ and $\tau$ for illustration. The initial state is set as $\alpha=0$ and two sets of parameters $J=0.0,U=0.5$ and $J=0.2,U=0.5$ are used ($J'=\pi/2+0.12,\Delta=0.1$ as before). Fig. \ref{FigureS3} shows the overlaps between time-evolved states obtained using different $\chi$ and $\tau$. The overlaps decrease as time increases because of the errors mentioned above. For a fixed $\tau$, the overlap between two neighboring $\chi$'s gradually increases. For the largest $\chi$, using $\tau=0.04,0.02,0.01$ give very similar results. This demonstrates that we can get excellent convergence using sufficiently large $\chi$ and small $\tau$. All the results in Figs. \ref{FigureS1} and \ref{FigureS2} were obtained using $\tau=0.01$ for several different $\chi$ values.

		\section{Experimental proposals}
		
		\subsection{Dipolar gases}
		In recent years, dipolar atoms (i.e. $ ^{168} $Er \cite{Er}, $ ^{160} $Dy, $ ^{161} $Dy, $ ^{162} $Dy \cite{dy1, dy2}) and molecules (i.e. $ ^{40} $K$ ^{87} $Rb \cite{krb}, $ ^{23} $Na$ ^{40} $K \cite{nak}, $ ^{23} $Na$ ^{87} $Rb \cite{narb}, $ ^{87} $Rb$ ^{133} $Cs \cite{rbcs}) have been widely prepared in cold atom experiments. Here we consider for example the fermionic molecule $ ^{23}$Na$ ^{40} $K prepared by Zwierlein's group at MIT, where the stable ground state of the Feshbach molecules has been achieved. Note that the ground state for $ ^{23} $Na$ ^{40} $K is a spin singlet, and therefore the fermionic nature of the dipolar molecule forbids double-occupancy of the same site due to Pauli blocking. Thus, there is no onsite interactions in addition to the dipolar one. (For bosonic molecules, one can similarly consider a deep optical lattice with strong onsite $ s $-wave repulsive interaction, such that double occupancy is also suppressed in such a hard-core regime. In the following we focus on fermionic particles and do not elaborate on the bosonic case).
		
		\subsubsection{Laser set-up}

		\begin{figure}
			[h]
			\parbox{8cm}{\begin{center}
					(a)
				\end{center} \quad\\
				\includegraphics[width=8cm]{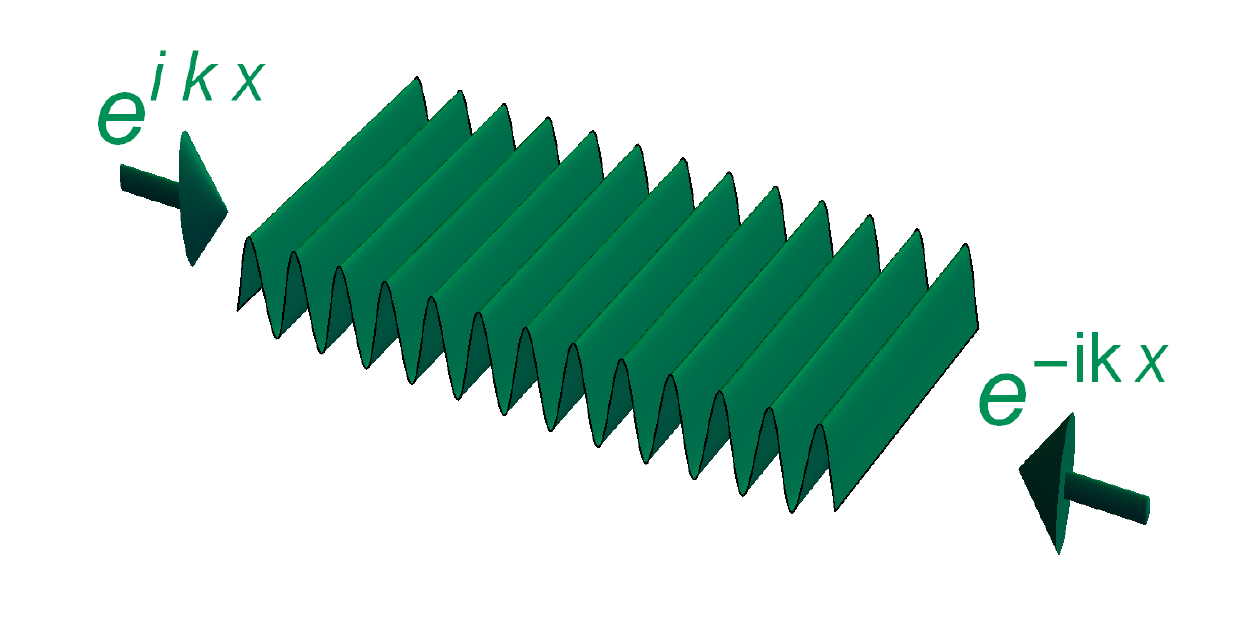}\\ \quad\\}
			\parbox{5cm}{
				\begin{center}
					(b)
				\end{center}\quad \\
				\includegraphics[width=6cm]{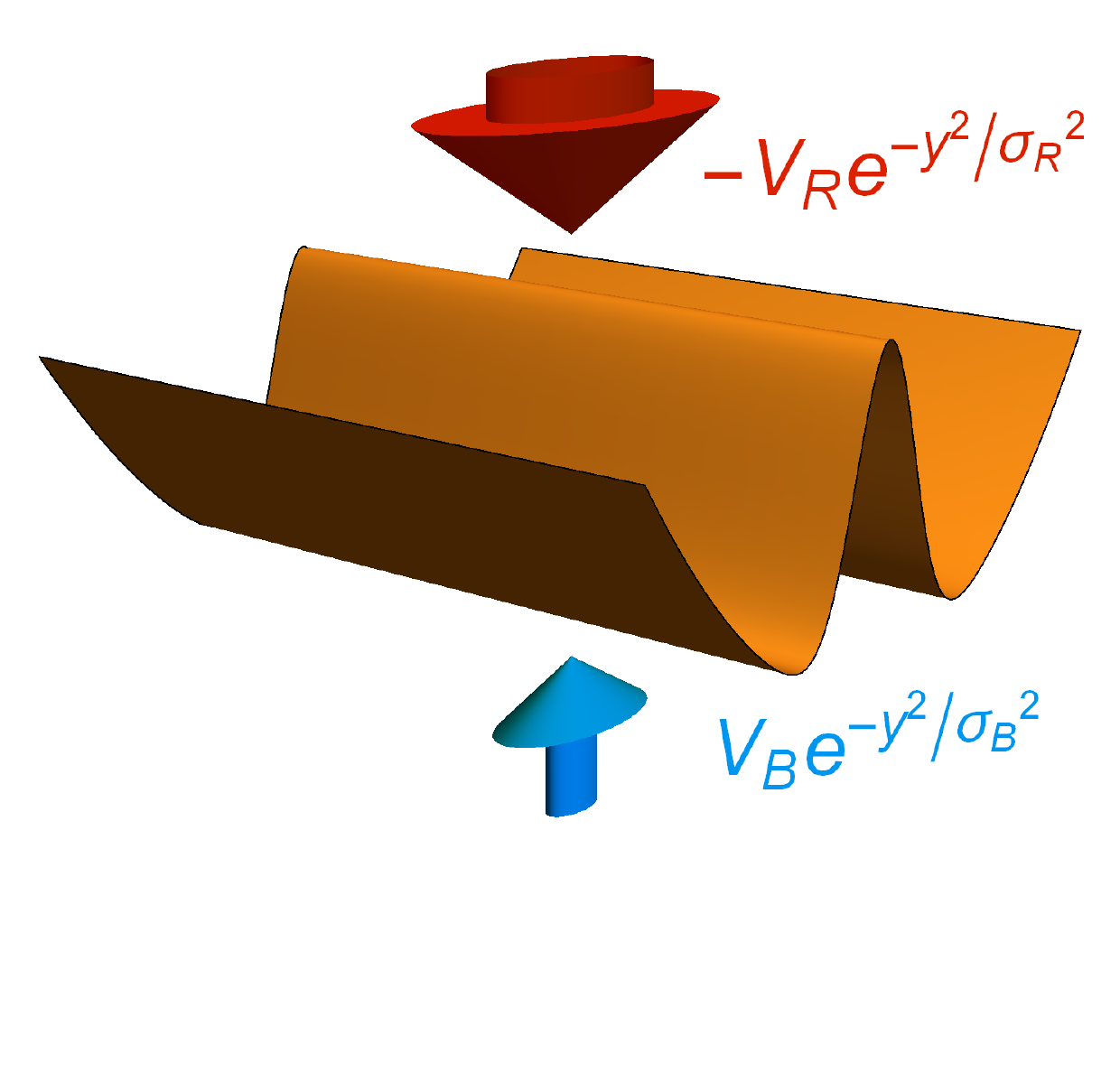}}
			\caption{\label{fig:dipolescheme}
				Schematic plot for the potentials engineered by superposing 4 laser beams. (a) A pair of laser beams along x-direction forms the 1D optical lattice. (b) Another two beams of lasers form the double-well potential along y-direction. Here the blue-detuned laser beam has much smaller Gaussian width and much stronger intensity than the red-detuned laser beam, so it gives the barrier for the double well, while the red-detuned laser gives the overall harmonic trap. Tuning the strength of blue-detuned beam $ V_B $ changes the barrier height, and tuning the width $ \sigma_B $ changes the distance between the two chains. The Gaussian width for both beams in x- and z-directions is much larger than either $ \sigma_R $ or $ \sigma_B $ and are not represented in the figure.}
		\end{figure}
		We propose using 2 beams of laser to form a double-well potential and to trap the gas in a quasi-1D system; and then using an additional pair of laser beams to form 1D optical lattices. The schematic plot is given in Fig.~\ref{fig:dipolescheme}. Note that the double well potential can be engineered in cold atom experiments in multiple ways, and here we adopt one of the schemes given in Ref.\cite{doublewellexpt}. Our modification is to increase the anisotropy for the beams in Fig.~\ref{fig:dipolescheme}(b) so a quasi-1D system is formed. The whole optical potential can be read as
		\begin{equation}\label{eq:dipole}
		V(x,y)=V_{latt}(t)\cos^2(kx) -V_R e^{-y^2/\sigma_R^2} + V_B(t) e^{-y^2/\sigma_B^2}
		\end{equation}
		The combined potential is given in the maintext by Fig.~\ref{fig:dipolescheme}. Here $ \sigma_R\gg \sigma_B $, so the red-detuned beam functions as an overall anisotropic harmonic trap for a quasi-1D system, while the blue-detuned beam gives the barrier between two wells. $ \sigma_B $ controls the distance between the two chains, and $ V_B $ tunes the height of the barrier and therefore the tunneling. The height and width can be tuned accurately within a wide range \cite{doublewellexpt}, providing feasible platform for the quench process. Also note that the initial state where particles fill up one of the two chains can be prepared by slowly moving the barrier from the edge to the center \cite{doublewellexpt}.
		
		\subsubsection{Dipolar interaction}
		
		\begin{figure}
			[h]
			\parbox{7.5cm}{\includegraphics[width=7cm]{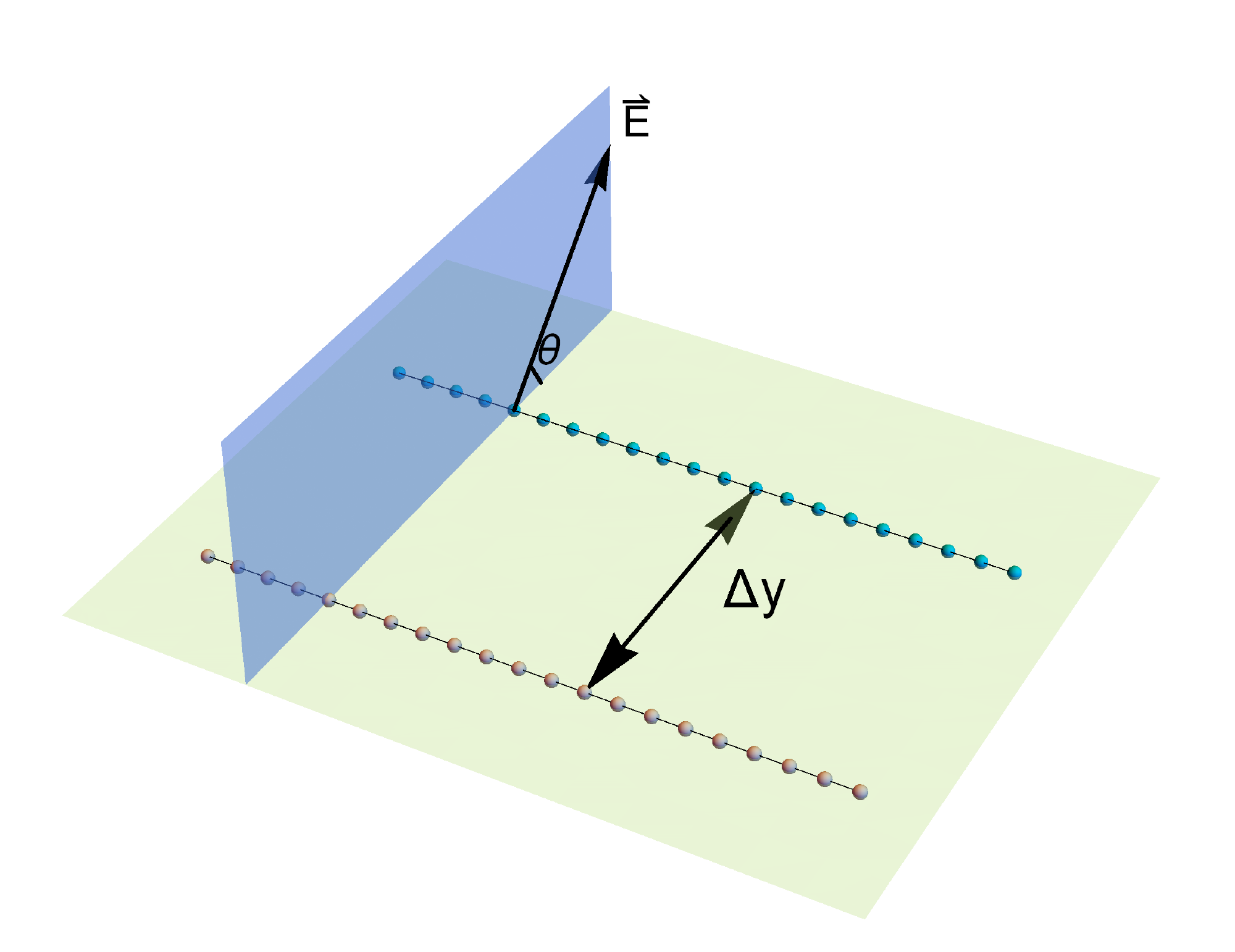}\\ \begin{center}
					(a)
				\end{center}}
				\parbox{7.5cm}{\includegraphics[width=7cm]{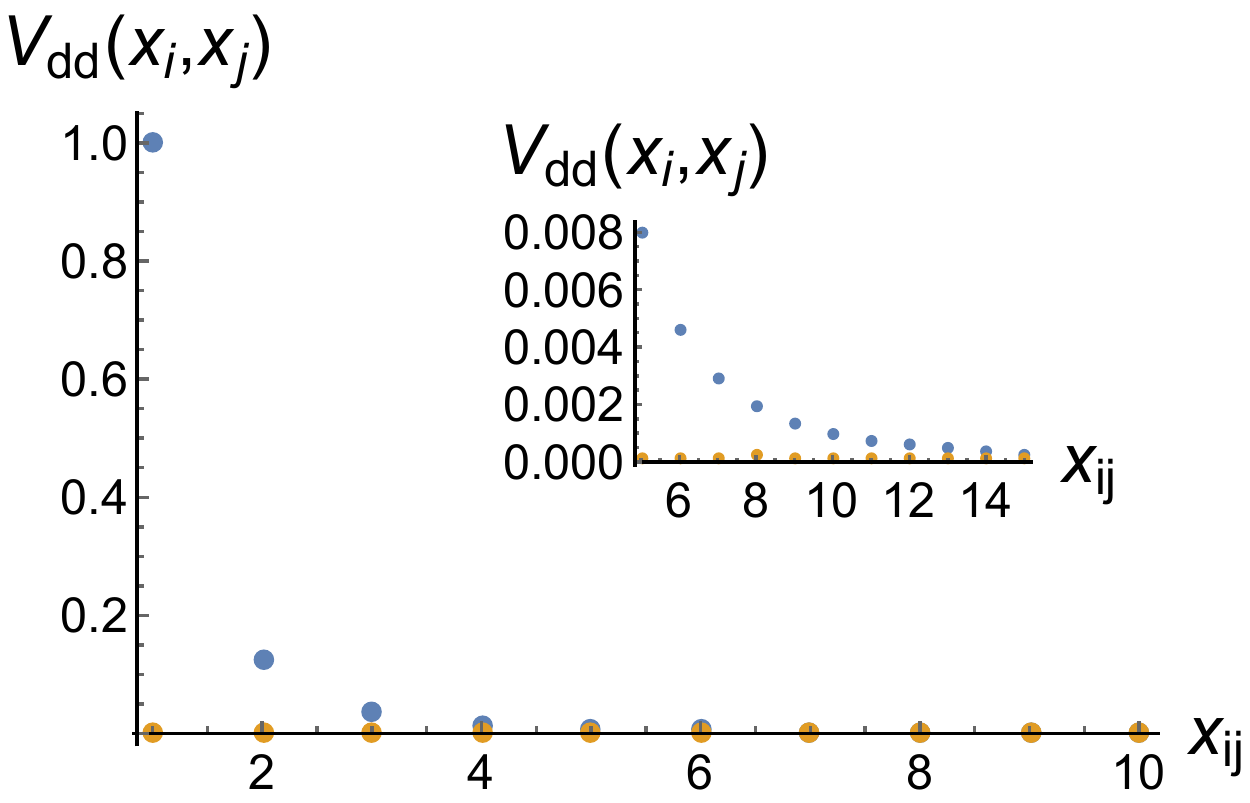}\\ \begin{center}
						(b)
					\end{center}}
					\caption{\label{fig:dipolar-polarization}
						(a) Schematic plot for the direction of electric field that polarizes the dipolar gas. $ \vec{E} $ is the electric field directions during $ t_2 $, given by Eq.~(\ref{E2}). The inter-chain spacing $ \Delta y $ is chosen to be large so as to reduce the dipolar interactions between two chains during $ t_2 $. (b) The interaction within each chain $ V^{A,A}_{dd} $ (blue dots) and between different chains $ V^{A,B}_{dd} $ (orange dots) during $ t_2 $. Although the inter-chain interaction is not strictly zero, it is vanishingly small. Here the units for interaction strength is $ U=d^2/4\pi\varepsilon_0 a^3 $, and the unit for site-distance $ x_{ij}=1,2,\dots $ along x-direction is the lattice spacing $ a $. We see that the interaction for sites in different chains is vanishing-small for the exemplary parameters we choose.}
				\end{figure}
				A general electric dipole interaction reads
				\begin{equation}\label{eq:generaldipole}
				V_{dd}(\mathbf{r}_i,\mathbf{r}_j) = \frac{d^2}{4\pi\varepsilon_0} 
				\frac{\hat{n}_i\cdot\hat{n}_j - 3(\hat{n}_i\cdot\hat{r}_{ij}) (\hat{n}_j\cdot\hat{r}_{ij})}{r_{ij}^3}
				\end{equation}
				Here $ d $ is the dipole strength, and $ \varepsilon_0$ is the dielectric constant in vacuum. $ \mathbf{r}_i, \mathbf{r}_j $ are the coordinate of the location for the two dipoles. $ r_{ij}=|\mathbf{r}_i-\mathbf{r}_j| $ is the distance for the two dipoles, and $ \hat{r}_{ij} = (\mathbf{r}_i-\mathbf{r}_j)/|\mathbf{r}_i-\mathbf{r}_j| $. The unit vector $ \hat{n}_i $ is the direction where the dipole is pointing to.
				
				For ultracold dipolar gases, the dipole direction $ \hat{n}_i $ is usually polarized by a uniform electric field, so $ \hat{n}_i\cdot\hat{n}_j=1 $ as $ \hat{n}_i\varparallel \vec{E} $. Then we can denote the site-independent electric field direction as $ \hat{n} $, and 
				\begin{equation}
				V_{dd}(\mathbf{r}_i,\mathbf{r}_j) = \frac{d^2}{4\pi\varepsilon_0}\frac{1-3(\hat{n}\cdot \hat{r}_{ij})^2}{r_{ij}^3}.
				\end{equation}

				During $ t_2 $, we consider electric field within the plane perpendicular to chain, forming an angle $ \theta $ with respect to the quasi-1D plane, as shown in Fig.~\ref{fig:dipolar-polarization}(a). As such, the interaction within each chain (i.e. within chain A) is
				\begin{equation}
				V_{dd}^{A,A}(i,j) = \frac{d^2}{4\pi\varepsilon_0 x_{ij}^3},\qquad x_{ij}=x_i-x_j.
				\end{equation}
				For interaction between two chains, 
				\begin{equation}\label{eq:dipole_interchain}
				V_{dd}^{AB}(i,j) = \frac{d^2}{4\pi\varepsilon_0 ((\Delta y)^2+x_{ij}^2)^{3/2}} \times \left(1-3\cos^2\theta\frac{(\Delta y)^2}{(\Delta y)^2+x_{ij}^2} \right)
				\end{equation}
				In general, the inter-chain interaction is never strictly zero: one can consider the limit where two sites separate far away in the x-direction, $ x_{ij}\rightarrow\infty $, then $ V_{dd}^{AB} (i,i+\infty)\rightarrow V_{dd}^{AA}(i,i+\infty) $. However, we note that the dipole interaction strength decays to the third power with respect to distance, and we can approximate a vanishing inter-chain interaction strength if the nearby sites have vanishing inter-chain interaction strength. Specifically, we choose a large $ \Delta y $, and let 
				\begin{equation}\label{E2}
				\cos^2\theta = \frac{1}{3},\qquad  \vec{E}_2\varparallel (0,1/\sqrt{3},\sqrt{2/3}).
				\end{equation}
				In Fig.~\ref{fig:dipolar-polarization}(b), we plot the interaction strength for intra-chain and inter-chain sites, with $ \Delta y= 10a $, where $ a $ is the lattice spacing. We clearly see that for small $ x_{ij} $, the interaction strength between different chains is vanishingly small. The large separation of two chains is also helpful for distinguishing particles in two chains in the imaging process.
				
				One may be worried whether the long-range part of the inter-chain interactions may play important roles. The concern derives from the experience in equilibrium systems, where the long-range tail of the interactions, albeit weak, can possibly change qualitative behaviors of the system. For instance, a Coulomb type of long-range interaction may induce a Wigner crystal behavior, while if one cuts off the interaction beyond certain distance, the phase may be qualitatively changed. However, note that here we are dealing with a highly non-equilibrium system subject to repeated quenches. The interaction does not last for infinitely long as in thermal equilibrium before the Hamiltonian is quenched. Therefore, we expect the weak long-range part of the inter-chain interaction exhibits negligible effects in our system. 

				\subsubsection{Parameter estimations and the quench process}
				
				There are four parameters in the model (Eq.~(3) and (7) in the main text): $ \theta, J $ for inter-chain and intra-chain tunneling, $ U $ for dipolar interaction strength, and $ \Delta $ for global chemical potential bias between two chains. These parameters also give the duration time for each quench period, and constrain how many periods can we have for one experiment. Here $ \theta, J, \Delta $ are all controlled by optical potential and can be tuned within a wide range in cold atom experiments. The chief constraint comes from dipolar interaction strength $ U $. Here we take the number from Ref.  \cite{nak}. Written in our notations,
				\begin{equation}
				U=d^2/4\pi\varepsilon_0a^3,
				\end{equation}
				where $ d\approx 0.3\sim0.9 $Debye is the dipole strength, 1Debye$ \approx 3.33\times 10^{-30} C\cdot m $; $ \varepsilon_0 = 8.85\times 10^{-12} F/m $ is the dielectric constant in vacuum; $ a $ is the lattice spacing within each chain. We set 
				\begin{equation}
				d=\alpha \times 1\mbox{Debye},\qquad
				a = \beta\times 1\mu m,
				\end{equation}
				then
				\begin{equation}
				U/\hbar \approx \frac{\alpha^2}{\beta^3} \times 950Hz,
				\end{equation}
				where $ \hbar=1.06\times 10^{-34} J\cdot s $ is the Planck constant. Take a typical number for experiments, $ \alpha=0.3$ ~\cite{nak}, $\beta=0.7 $~\cite{pethicksmith}, we have $ U/\hbar\approx 250 $Hz. On the other hand, the typical lifetime for the dipolar system is a few seconds~\cite{nak}. Thus, we estimate that a typical experiment can undergo several hundreds of Floquet periods.
				
				During $ t_1 $, $ V_{latt} $ in Eq.~(\ref{eq:dipole}) takes large value and the barrier $ V_B $ is lowered, so there is effectively only tunneling between the rungs of the ladder. As discussed before, the electric field is along the $ (1/\sqrt{3})(1,1,1) $ direction. Then the dipole interaction within the same chain is strictly zero, while interaction between different chains are smaller than $ 0.6\times 10^{-3} U$. Note the tunneling strength $ \theta/h $ is of the order of $ 10^{2} $Hz.\footnote{The order of magnitude of tunneling amplitude can be estimated by multiplying the overlap of Wannier wavefunctions times the recoil energy $ E_r=\hbar^2 k^2/2m $, where $ k $ is the wave vector for lasers, $ m $ is the mass for molecules. For $ ^{23} $Na$ ^{40} $K, $ m\approx 10^{-25} $kg, and the laser wavelength is of the order of $ \lambda \sim 0.5\mu m $; then, $ E_r/h $ is of the order of $ 10^{3\sim 4} $Hz. The tunneling strength goes down exponentially as one ramps up the lattice depth so completely isolated lattice sites can be easily achieved (For dipolar molecules in optical lattices, see i.e.  \cite{krb2})} Thus, during $ t_1 $ we can safely neglect dipole interactions and achieve $ H_1 $ given by Eq.~(2) in the main text. The duration $ t_1 $ is given by $ (\pi/2+\varepsilon)\hbar/\theta $, which is typically tens of miliseconds.

				During $ t_2 $, one ramps up the inter-chain barrier $ V_B $ and turns down the 1D lattice potential $ V_{latt} $. Also, one turns down the electric field so that the dipole moment goes down. Then each chain separately undergoes intra-chain tunneling $ J $ as well as dipolar interactions. The duration can be estimated by $ \hbar/U $.

				\subsection{SU(N) particles}
				The set-up for SU(N) particles shares many similarities with that for dipolar gases. And in many aspects, it may to be even simpler to implement in experiments.
				\subsubsection{Laser set-up}	
				For SU(N) particles, each ``chain'' is represented by different internal states, i.e. the spin $ m_F=-S, -S+1, \dots, S $, and $ N=2S+1 $. Thus, we only need a double-well potential to represent the ``rung'' direction, i.e. removing the $ V_{latt} $ in Eq.~(\ref{eq:dipole}). The initial state consists of $ (2S+1) $ particles confined within one of the two wells.
				
				In principle, one can use Raman coupling of different spin states to engineer the ``tunneling'' $ J $ within each chain along the synthetic dimension. Moreover, the tunneling term $ J $ can be a complex number, which may lead to interesting phenomenon, such as quantum-Hall like physics \cite{syntheticD1,syntheticD2}, due to the non-zero flux within a plaquettee of the ladder. We leave the discussion for Raman-coupled spin states for future work, and here only focus on the $ J=0 $ situation, which is easier to implement in experiments.
				
				\subsubsection{Parameter estimations}
				The key feature of SU(N) particles is that due to the lack of electronic spin (so the hyperfine spin equals the nuclear spin), the collisions between SU(N) particles does not flip the hyperfine spin state. Thus, the hyperfine magnetic quantum number $ m_F $ functions as a good quantum number representing the (spin-)site index. We can estimate the interaction strength $ U/h $ in our model by noting that the scattering lengths  \cite{sunSc1, sr1, sr2, yb1, yb2}
				\begin{equation}\label{sunscattering}
				\mbox{$ ^{173} $Yb: }a_s = 10.55 nm,\qquad\qquad
				\mbox{$ ^{87} $Sr: }a_s = 5.05nm.
				\end{equation}
				The interaction strength
				\begin{equation}
				U= \frac{4\pi\hbar^2 a_s}{m}\int d^3x |\psi(\mathbf{x})|^4,
				\end{equation}
				where $ m $ is the mass for SU(N) particles, and $ \psi(\mathbf{x}) $ is the wave function for an atom in one of the double well potential. Note that both the scattering lengths and the mass for $ ^{173} $Yb are about twice as much as those for $ ^{87} $Sr; therefore, the interaction strengths $ U $ for these two types of atoms are actually the same, depending only on the size of the wave function. Note that as an estimation, we can approximate $ \int d^3x |\psi (\mathbf{x})|^4 \approx 1/L^3 $, where $ L $ is the size of the wave function that can be tuned by changing the depth and size of the double-well potential.  Consider, for example, that $ L\approx 2\mu m $, we have 
				\begin{equation}
				\frac{U}{\hbar}\approx 5.8 Hz.
				\end{equation}
				Then for $ Ut_2/h=0.1 $, we have $ t_2\approx 17.2 $ miliseconds. Note that the size of the well (a few micrometers) we consider is much larger than the scattering lengths (\ref{sunscattering}). So it is not in a ``deep optical lattice'' regime where atom loss may be severe due to 3-atom recombinations  \cite{sunSc1,sr1}. Thus, we expect the life-time for the system is the same as a typical SU(N) fermion system, which can last for a few seconds. That implies that a few hundreds of Floquet periods can be observed. 
				
				Finally, we note that during $ t_1 $, as estimated in the case of dipolar gases, the tunneling strength $ \theta/h $ is of the order of $ 10^{2} $Hz, which is several orders of magnitudes larger than the interaction strength $ U/h $. Thus, we can treat the system in this regime as non-interacting particles, and end up with $ H_1 $ in Eq.~(2) in the main text. (In principle, one can further apply a magnetic field to reduce the interaction strength  \cite{pethicksmith}).

\end{document}